\documentclass[aps,prd,onecolumn,nofootinbib,superscriptaddress]{revtex4}
\usepackage{graphicx}
\usepackage{subfigure}
\usepackage{amsmath}
\usepackage{amsfonts}
\usepackage{amssymb,ulem}
\usepackage{color}%
\usepackage{booktabs}
\usepackage{amsmath}
\usepackage{dcolumn}

\usepackage{MnSymbol,wasysym}
\usepackage{braket}
\usepackage{eurosym}
\usepackage{calrsfs}
\usepackage{multirow}
\usepackage{float}

\newcommand{\RNum}[1]{\uppercase\expandafter{\romannumeral #1\relax}}
\usepackage[colorlinks=true,linkcolor=blue,urlcolor=blue,citecolor=red]{hyperref}
\usepackage[title]{appendix}
\makeatletter
\newcommand*{\rom}[1]{\expandafter\@slowromancap\romannumeral #1@}
\makeatother

\begin{document}
\baselineskip=0.5 cm

\title{Periodic Timelike Motion and Gravitational Wave Signatures around a Magnetically Charged Black Hole Surrounded by Quintessence}

\author{R. H. Ali}
\email{hasnainali408@yzu.edu.cn, hasnainali408@gmail.com}
\affiliation{Center for Gravitation and Cosmology, College of Physical Science and Technology, Yangzhou University, Yangzhou, 225009, China}

\author{M. R. Shahzad}
\email{dr.rizwanshahzad@bzu.edu.pk}
\affiliation{Institute for Theoretical Physics and Cosmology, Zhejiang University of Technology, Hangzhou 310023, China}
\affiliation{United Center for Gravitational Wave Physics (UCGWP), Zhejiang University of Technology, Hangzhou, 310023, China}
\affiliation{Department of Mathematics, Bahauddin Zakariya University, Vehari Campus, Vehari 61100, Pakistan.}
\affiliation{Research center of Astrophysics and Cosmology, Khazar University, Baku, AZ1096, 41 Mehseti Street, Azerbaijan.}

\author{Hamza Rehman}
\email{hamzarehman244@zjut.edu.cn}
\affiliation{Institute for Theoretical Physics and Cosmology, Zhejiang University of Technology, Hangzhou 310023, China}
\affiliation{United Center for Gravitational Wave Physics (UCGWP), Zhejiang University of Technology, Hangzhou, 310023, China}

\author{Ahdab K. Althukair}
\email{akalthukair@pnu.edu.sa}
\affiliation{Department of Physics, College of Sciences, Princess Nourah bint Abdulrahman University.}

\author{Walid Abdelfattah}
\email{walid.abdelfattah@nbu.edu.sa}
\affiliation{Department of Mathematics, Northern Border University, Arar, Saudi Arabia}

\begin{abstract}
\baselineskip=0.5 cm
We investigate timelike geodesics and gravitational wave signatures of periodic motion around a static magnetically charged black hole arising from nonlinear electrodynamics and immersed in a quintessence background. We analyze the effective potential for massive particles, determine the marginally bound and innermost stable circular orbits, and classify the resulting bound trajectories using the zoom-whirl taxonomy $(\mathit{z},\mathit{w},\mathit{v})$. We show that the quintessence parameter $c_q$ systematically shifts the orbital radii, conserved quantities, and turning-point structure associated with representative periodic families. We then model the gravitational radiation emitted by periodic extreme-mass-ratio inspirals within the numerical kludge approximation. The resulting waveforms exhibit the characteristic burst-like structure of zoom-whirl motion, while variations in the quintessence coupling parameter modify the phase evolution, burst timing, and harmonic content of the signal. The corresponding Fourier spectra display a discrete comb-like structure, and the characteristic strain is concentrated in the millihertz band relevant for space-based detectors such as LISA. These results indicate that a quintessence background can leave systematic imprints on periodic orbit dynamics and on the associated time and frequency-domain gravitational wave observables.
\end{abstract}

\maketitle
\newpage
\tableofcontents

\section{Introduction}
Black holes (BHs) are exact solutions of Einstein's field equations in general relativity (GR) and constitute one of the theory's most remarkable predictions. Nearly a century after their original theoretical formulation, they have evolved from purely mathematical constructs into indispensable laboratories for exploring gravity in the strong-field regime. Among the known solutions, the Schwarzschild and Kerr metrics~\cite{Blinder:2015ecg,Kerr:1963ud} represent the canonical vacuum geometries of Einstein's equations and form the basis of much of modern BH physics. Recent observational advances have elevated BHs to the forefront of strong-gravity research. The direct detection of gravitational waves (GWs) from compact-binary coalescences by the LIGO and Virgo collaborations~\cite{LIGOScientific:2016aoc,LIGOScientific:2018mvr,LIGOScientific:2020aai}, the horizon-scale imaging of BH shadows by the Event Horizon Telescope~\cite{EventHorizonTelescope:2019dse,EventHorizonTelescope:2019ths,EventHorizonTelescope:2019pgp,EventHorizonTelescope:2022wkp,EventHorizonTelescope:2022xqj}, and the precise monitoring of stellar orbits around the compact object $\mathrm{Sgr~A^*}$ at the Galactic center together provide compelling evidence for the existence of astrophysical BHs \cite{Genzel:2010zy,Iorio:2011zi, Grould:2017bsw, DeLaurentis:2018ahr, GRAVITY:2019tuf,DES:2019ltu}. At the same time, these observations furnish increasingly stringent tests of GR in the nonlinear regime that may encode signatures of new gravitational physics.

In this context, the study of particle motion in BH spacetimes provides one of the most direct ways to probe the geometry of strong gravitational fields. Geodesic trajectories of massless and massive particles encode a wide range of observable phenomena, including BH shadows, gravitational lensing, and orbital dynamics near compact objects. In particular, null geodesics determine the photon capture region and hence the morphology of BH shadows, whereas timelike geodesics govern the motion of massive particles and therefore carry direct information about the dynamics of stars and compact bodies in the vicinity of astrophysical BHs. As a result, geodesic analysis has become a central tool for extracting physical information from BH spacetimes and for assessing how modifications of the underlying geometry manifest themselves in observable strong-field phenomena.

Among timelike trajectories, periodic orbits are of special interest because they provide a refined characterization of the orbital structure of the spacetime and are especially sensitive to the near-horizon geometry. In the strong-field regime, the properties of bound and periodic timelike motion encode nontrivial information about the effective gravitational potential, orbital stability, and the transition between regular and highly relativistic motion. These features are particularly relevant in the context of GW astronomy. The first direct detection of GWs by the LIGO and Virgo collaborations opened an unprecedented observational window onto the dynamics of compact objects in strong gravity~\cite{LIGOScientific:2016vbw,LIGOScientific:2016vlm,LIGOScientific:2016emj}. Since GW signals are generated by the orbital motion of compact bodies in curved spacetime, they provide direct access to the dynamical structure of the background geometry. It is therefore natural to expect that periodic timelike orbits, as probes of the strong-field orbital sector, can furnish valuable information about the GW signatures associated with nontrivial BH spacetimes.

Beyond their role in transient strong-field phenomena, BH spacetimes also provide a natural setting for exploring the dynamical structure of gravity through the motion of test particles. Within this framework, periodic orbits occupy a distinguished position, since they form the backbone of the orbital phase space and encode key information about the underlying nonlinear dynamics. Their properties are closely tied to orbital stability, resonant behavior, and the global organization of geodesic motion in curved spacetime~\cite{Levin:2008mq,Levin:2009sk,Misra:2010pu,Babar:2017gsg}. From a dynamical perspective, generic bound trajectories may be viewed as deformations of periodic orbits, which therefore provide a natural basis for understanding the structure of bound motion in strong gravitational fields. This viewpoint has proved particularly useful for analyzing the motion of test particles and accreting matter in BH backgrounds, and it is especially relevant for extreme-mass-ratio inspirals (EMRIs), where a stellar-mass compact object gradually inspirals into a supermassive BH under the backreaction of gravitational radiation.

EMRIs are among the most promising targets for next-generation space-based GW observatories, including LISA~\cite{Danzmann:1997hm,Schutz:1999xj,Gair:2004iv,LISA:2017pwj,Maselli:2021men}, TianQin~\cite{TianQin:2015yph,Gong:2021gvw}, and Taiji~\cite{Hu:2017mde}. Because these systems probe the deep strong-field region over a large number of orbital cycles, they offer a unique opportunity to test the near-horizon structure of astrophysical BHs with high precision. In this context, the classification of periodic timelike trajectories provides a systematic framework for organizing the orbital dynamics and identifying the characteristic features of strong-field motion. Such a program has been pursued extensively in a wide range of BH geometries, including Schwarzschild, Kerr, charged, regular, and hairy spacetimes~\cite{Grossman:2011im,Tu:2023xab,Li:2024tld,Deng:2020yfm,Lin:2021noq,Gao:2021arw,Zhang:2022psr,Lin:2022wda,Habibina:2022ztd,Wang:2022tfo,Yao:2023ziq,Lin:2023eyd,
Chan:2025ocy,Haroon:2025rzx,Wang:2025wob,Alloqulov:2025bxh,Wei:2025qlh,Azreg-Ainou:2020bfl}, and the associated GW signatures of periodic motion have likewise been investigated in several BH backgrounds~\cite{Junior:2024tmi,Yang:2024lmj,Zhao:2024exh,Jiang:2024cpe,Yang:2024cnd,Meng:2024cnq,QiQi:2024dwc,Shabbir:2025kqh,Alloqulov:2025ucf,Wang:2025hla,Lu:2025cxx,
Zare:2025aek,Gong:2025mne,Li:2025sfe,Choudhury:2025qsh,Chen:2025aqh,Deng:2025wzz,Li:2025eln,Zahra:2025tdo,Yang:2026syx}. These developments provide strong motivation for extending the analysis of periodic orbits and their GW imprints to other nontrivial BH spacetimes.

Black holes sourced by nonlinear electrodynamics (NED) constitute a well-motivated extension of the standard Einstein-Maxwell family, in which nonlinear electromagnetic self-interactions deform the spacetime geometry and can generate charged solutions with nontrivial strong-field structure~\cite{Ghosh:2021clx,Kruglov:2020tes,Bazrafshan:2019ege,Panotopoulos:2018rjx,Toshmatov:2017zpr,Flores-Alfonso:2020euz,Huang:2025vpi,Bronnikov:2024izh,Aydiner:2025eii,
Hazarika:2025axz,Azreg-Ainou:2025tuj,Jha:2025cqf,Saikia:2025sfj,Liang:2025jph,Sucu:2024xck}. Owing to the sensitivity of strong-field orbital dynamics to the underlying background geometry, such NED induced deformations provide a natural arena for investigating timelike geodesics, periodic orbits, and their associated GW signatures in EMRI scenarios~\cite{Zahra:2025tdo,Alloqulov:2025bxh,Figliolia:2026qys}; more generally, modifications of the spacetime geometry can alter the orbital motion of inspiralling compact objects and thereby affect the corresponding GW emission~\cite{Mkrtchyan:2022ulc,Bisnovatyi-Kogan:2017kii,Atamurotov:2021imh}. In parallel, BHs surrounded by a quintessence field have been widely studied as an effective phenomenological framework for incorporating dark-energy like matter into spherically symmetric spacetimes. In the Kiselev construction, the quintessential medium is modeled by an anisotropic fluid characterized by an equation of state parameter $\omega_q$, which introduces an additional deformation of the metric and can substantially modify the horizon structure, effective potential, and orbital properties of the spacetime~\cite{Kiselev:2002dx,Uniyal:2014paa,Maurya:2025jch,Gohain:2024aod,Gohain:2024piy,Ovchinnikov:2021xfw,Dariescu:2026vkh,Rayimbaev:2024ywi,Rayimbaev:2022mrk}. The influence of quintessence on periodic orbital dynamics and the associated GW phenomenology has likewise attracted increasing attention in recent years~\cite{Wang:2022tfo,Li:2025eln,Ahmed:2025azu}.

Taken together, these developments strongly motivate the study of BH geometries in which NED effects and a quintessence background are simultaneously present, since the interplay between magnetic charge, NED corrections, and quintessential matter can leave nontrivial imprints on the structure of timelike periodic motion and, consequently, on the GW signals emitted by orbiting compact objects. Motivated by this framework, we consider a static magnetically charged BH immersed in a quintessence background, obtained in Einstein gravity coupled to a Power-Maxwell NED sector together with a Kiselev-type quintessence source. The dynamics of the model are governed by the action~\cite{Kiselev:2002dx,Hasnain:2025alz}

\begin{equation}
\mathbf{S}=\int d^4x\,\sqrt{-g}\left[\frac{R}{16\pi}+\mathcal{L}_{\rm EM}+\mathcal{L}_{q}\right],
\end{equation}

where $R$ is the Ricci scalar, $\mathcal{L}_{\rm EM}$ denotes the Lagrangian density of the Power-Maxwell NED sector, and $\mathcal{L}_{q}$ represents the quintessential contribution introduced phenomenologically through its energy-momentum tensor following the Kiselev prescription~\cite{Kiselev:2002dx}. The electromagnetic Lagrangian is taken as
\begin{equation}
\mathcal{L}_{\rm EM}=\beta(-\mathcal{F})^{s},
\end{equation}
where $\mathcal{F}\equiv \mathcal{F}_{\mu\nu}\mathcal{F}^{\mu\nu}$, $s$ is a real parameter characterizing the nonlinearity of the electromagnetic sector, and $\beta$ is the corresponding coupling constant. In this geometry, the magnetic charge contributes a charge dependent correction to the metric function, while the quintessence field introduces an additional deformation governed by the equation of state parameter $\omega_q$. As a result, the combined influence of NED and quintessential matter modifies the effective potential governing particle motion and can therefore shift the location of circular orbits, alter stability regions, and reshape the structure of periodic timelike trajectories.

In this paper, we investigate the GW emission generated by the periodic timelike motion of a test particle around this magnetically charged BH in a quintessence background. In particular, we examine how the quintessence coupling parameter affects the orbital dynamics, classify the resulting periodic trajectories, and compute the corresponding GW radiation. Through this analysis, we seek to identify how NED and quintessence-induced modifications of the background geometry are encoded in both the periodic orbit structure and the emitted waveform.

The paper is organized as follows. In Sec.~\ref{sec:BH-review and geodesics}, we briefly review the magnetically charged BH solution in the presence of NED and construct the timelike geodesic equations, with particular emphasis on how the model parameters affect the orbital dynamics and the locations of the innermost stable circular orbit (ISCO) and marginally bound orbit (MBO). In Sec.~\ref{sec:Periodic Orbits}, we analyze the structure and classification of periodic timelike orbits in this background. In Sec.~\ref{sec:GW}, we study the GW signals emitted by periodic orbits in the corresponding EMRI system. Finally, we summarize our main results and present concluding remarks in Sec.~\ref{sec:con}.

\section{Black hole background and timelike geodesics}\label{sec:BH-review and geodesics}
Following Refs.~\cite{Kiselev:2002dx,Hasnain:2025alz}, we consider a static, spherically symmetric magnetically charged BH immersed in a quintessence background. The spacetime arises from Einstein gravity coupled to a Power-Maxwell NED sector together with a Kiselev-type quintessential source, and is described by the line element
\begin{equation}
ds^{2}=-N(r),dt^{2}+\frac{dr^{2}}{N(r)}+r^{2}\left(d\theta^{2}+\sin^{2}\theta,d\phi^{2}\right),\label{Eq:a0}
\end{equation}
with metric function
\begin{equation}
N(r)=1-\frac{2M}{r}+\frac{Q^{2}}{r^{2}}-\frac{c_{q}}{r^{3\omega_{q}+1}}, \qquad Q^{2}=2\pi Q_{m}^{2}. \label{Eq:a1}
\end{equation}
Here $M$ denotes the BH mass, $Q_m$ is the magnetic charge parameter, and $Q$ is the effective charge entering the metric function through the Einstein-Power-Maxwell sector. The parameter $\omega_q$ is the equation of state parameter of the surrounding quintessence field, while $c_q$ controls the amplitude of the quintessential contribution.
In the present work, we restrict attention to the parameter ranges
\begin{equation}
0\leq \frac{Q_m}{M}\leq 0.9, \qquad \omega_q\in[-0.7,-0.3],\label{Eq:paramrange}
\end{equation}
which keep the magnetic charge away from the near-extremal regime and place the quintessential component within the standard Kiselev-type domain.
The metric function in Eq.~\eqref{Eq:a1} contains three distinct contributions: the Schwarzschild mass term $-2M/r$, the magnetic charge term $Q^{2}/r^{2}$ inherited from the NED sector, and the quintessence contribution $-c_q/r^{3\omega_q+1}$. In the Kiselev construction, $c_q$ characterizes the strength of the surrounding quintessential background, so that $c_q=0$ recovers the charged vacuum geometry, whereas $c_q\neq0$ encodes the backreaction of the ambient dark-energy like medium. Through its direct contribution to the radial potential, the quintessence term modifies the horizon structure, the effective potential, and the location and stability of circular timelike orbits, and hence directly affects the existence and properties of periodic trajectories relevant for GW emission.

Equation~\eqref{Eq:a1} also contains several important limiting cases. For $Q_m=c_q=0$, it reduces to the Schwarzschild BH solution, while $c_q=0$ yields the RN geometry with the effective charge identification $Q^{2}=2\pi Q_m^{2}$. In the complementary limit $Q_m=0$, one recovers the Kiselev BH surrounded by quintessence. The present spacetime therefore provides a unified framework for analyzing the combined effects of magnetic charge and quintessential matter on strong-field orbital dynamics.

To analyze the motion of a massive test particle in the spacetime~\eqref{Eq:a0}, we consider timelike geodesics confined, without loss of generality, to the equatorial plane $\theta=\pi/2$. The dynamics of the particle are then governed by the Lagrangian~\cite{Chandrasekhar:1984siy} as
\begin{equation}
2\mathcal{L}=g_{\mu \nu}dx^{\mu}dx^{\nu}=-N(r)\dot{t}^{2}+\frac{1}{N(r)}\dot{r}^{2}+r^{2}\dot{\phi}^{2},\label{Eq:specific-metric}
\end{equation}
where the overdot represents the derivative against the affine parameter $\tau$.\\

The Euler-Lagrangian equation is of the form~\cite{Chandrasekhar:1984siy}
\begin{eqnarray}
&&p_{\mu}=\frac{\partial \mathcal{L}}{\partial \dot{x}^\mu}=g_{\mu \nu}\dot{x}^\nu=\begin{cases}
p_{t}=\frac{\partial \mathcal{L}}{\partial \dot{t}}=-N(r)\dot{t}=-\mathit{E}= const,\\
p_{\phi}=\frac{\partial \mathcal{L}}{\partial \dot{\phi}}=r^{2}\dot{\phi} =\mathit{L}= const,\\
p_{r}=\frac{\partial \mathcal{L}}{\partial \dot{r}}=\frac{\dot{r}}{N(r)},
\end{cases}\label{Eq:momenta}
\end{eqnarray}
where $\mathit{E}$ and $\mathit{L}$ typically express the conserved specific energy and specific angular momentum per unit mass of the particle.\\

Imposing the normalization condition $g_{\mu\nu}\dot{x}^{\mu}\dot{x}^{\nu}=-1$ for timelike geodesics, the radial equation of motion can be written as
\begin{equation}
\dot{r}^2=\mathit{E}^{2}-N(r)\Big(1+\frac{\mathit{L}^{2}}{r^{2}}\Big),\label{Eq:radial equation}
\end{equation}
from which the corresponding effective potential is identified as $V_{\rm eff}(r)$ as
\begin{equation}
V_{\rm eff}(r)\equiv \mathit{E}^{2}-\dot{r}^2=\Big(1-\frac{2M}{r}+\frac{Q^{2}}{r^{2}}-\frac{c_{q}}{r^{3\omega_{q}+1}}  \Big)\Big(1+\frac{\mathit{L}^{2}}{r^2}\Big).\label{Eq:effective-potential for timelike}
\end{equation}

The effective potential $V_{\rm eff}(r)$ depends on the conserved angular momentum $\mathit{L}$, the radial coordinate $r$, and the BH parameters entering the metric function. Since $V_{\rm eff}(r)\to 1$ as $r\to\infty$, the value $\mathit{E}=1$ marks the threshold between bound and unbound timelike motion. In particular, particles with $\mathit{E}>1$ can escape to infinity and correspond to scattering trajectories, whereas bound orbits satisfy $\mathit{E}\leq 1$ \cite{straumann2012general}. The interplay between $\mathit{E}$ and $V_{\rm eff}(r)$ therefore determines the qualitative structure of particle motion in spacetime.

In the present work, we focus exclusively on bound timelike motion, for which the particle energy and angular momentum must lie within a restricted region of parameter space. Our main interest is in periodic orbits, which form a distinguished subclass of bound trajectories in the BH spacetime. Before turning to the analysis of generic periodic motion, it is useful to identify two special limiting orbits that bound the relevant parameter domain: the MBO and the ISCO~\cite{misner1973gravitation}.

The existence of periodic orbits requires the conserved energy and angular momentum of the particle to satisfy~\cite{Dadhich:2021bzw,Qi:2024tim}
\begin{equation}
\mathit{L}_{\rm ISCO}\leq \mathit{L}, \qquad \mathit{E}_{\rm ISCO}\leq \mathit{E}\leq \mathit{E}_{\rm MBO}=1,
\label{Eq:specific condition}
\end{equation}
where $\mathit{L}_{\rm ISCO}$ and $\mathit{E}_{\rm ISCO}$ denote the angular momentum and energy of the particle at the ISCO, respectively, while $\mathit{E}_{\rm MBO}$ is the corresponding energy at the MBO.

We now turn to two special circular timelike orbits that delimit the parameter region relevant for periodic motion, namely the MBO and the ISCO. These orbits play a central role in the analysis of bound motion, since they determine the range of conserved energy and angular momentum for which periodic trajectories can exist in the BH spacetime.\\
The MBO is defined as the circular orbit at the threshold between bound and unbound motion and is therefore characterized by unit-specific energy.
Its radius is determined by the conditions
\begin{equation}
V_{\rm eff}(r)=1, \qquad \partial_r V_{\rm eff}(r)=0. \label{Eq:condition1}
\end{equation}
These equations are solved simultaneously to obtain the marginally bound radius $r_{\rm MBO}$ and the corresponding angular momentum
$\mathit{L}_{\rm MBO}$.\\

Another orbit of particular importance is the ISCO, which marks the transition between stable and unstable circular timelike motion and therefore sets the inner edge of the stable bound-orbit region. In the present spacetime, the ISCO is determined by the conditions
\begin{equation}
V_{\rm eff}(r)=E^{2}, \qquad \partial_r V_{\rm eff}(r)=0, \qquad \partial_r^{2}V_{\rm eff}(r)=0, \label{Eq:condition2}
\end{equation}
from which one extracts the ISCO radius together with the corresponding energy and angular momentum. Since the MBO and ISCO bound the parameter domain of periodic trajectories, their dependence on the magnetic charge and quintessence parameters provides direct information about how the background geometry influences the structure of periodic orbits and the associated GW phenomenology.

\begin{table}[htbp]
\centering
\caption{The values of $r_{\rm MBO}$, $r_{\rm ISCO}$, $\mathit{L}_{\rm ISCO}$ and $\mathit{L}_{\rm ISCO}$ with fixed value of charged parameter $Q$, and various values of $c_{q}$. }
\label{tab:ISCO_IBCO_lambda}
\renewcommand{\arraystretch}{1.2}
\begin{tabular}{|c|c|c|c|c|}
\hline
\multicolumn{5}{|c|}{$Q = 0.5$, $\omega_{q}=-0.30$} \\
\hline
~~~$c_{q}$~~~~~~~~~~~~~~~ & $r_{\rm MBO}$~~~~~~~~~~~~~~~ & $r_{\rm ISCO}$~~~~~~~~~~~~~~~ & $\mathit{L}_{\rm ISCO}$~~~~~~~~~~~~~~~ & $\mathit{L}_{\rm MBO}$\\
\hline
~~~0~~~~~~~~~~~~~~~    & 3.736872~~~~~~~~~~~~~~~ & 5.606643~~~~~~~~~~~~~~~ & 3.337737~~~~~~~~~~~~~~~ & 3.736872 \\ \hline

~~~0.01~~~~~~~~~~~~~~~ & 3.748801~~~~~~~~~~~~~~~ & 5.659522~~~~~~~~~~~~~~~ & 3.372844~~~~~~~~~~~~~~~ & 3.937039 \\ \hline

~~~0.02~~~~~~~~~~~~~~~ & 3.762270~~~~~~~~~~~~~~~ & 5.713250~~~~~~~~~~~~~~~ & 3.408586~~~~~~~~~~~~~~~ & 4.006850 \\ \hline

~~~0.03~~~~~~~~~~~~~~~ & 3.777187~~~~~~~~~~~~~~~ & 5.767850~~~~~~~~~~~~~~~ & 3.444981~~~~~~~~~~~~~~~ & 4.077997  \\ \hline

~~~0.04~~~~~~~~~~~~~~~ & 3.793479~~~~~~~~~~~~~~~ & 5.823340~~~~~~~~~~~~~~~ & 3.482045~~~~~~~~~~~~~~~ & 4.150527  \\ \hline

~~~0.05~~~~~~~~~~~~~~~ & 3.811082~~~~~~~~~~~~~~~ & 5.879744~~~~~~~~~~~~~~~ & 3.519795~~~~~~~~~~~~~~~ & 4.224490  \\ \hline

~~~0.06~~~~~~~~~~~~~~~ & 3.829943~~~~~~~~~~~~~~~ & 5.937083~~~~~~~~~~~~~~~ & 3.558250~~~~~~~~~~~~~~~ & 4.299936  \\ \hline

~~~0.07~~~~~~~~~~~~~~~ & 3.850017~~~~~~~~~~~~~~~ & 5.995380~~~~~~~~~~~~~~~ & 3.597429~~~~~~~~~~~~~~~ & 4.376916  \\ \hline

~~~0.08~~~~~~~~~~~~~~~ & 3.871267~~~~~~~~~~~~~~~ & 6.054660~~~~~~~~~~~~~~~ & 3.637350~~~~~~~~~~~~~~~ & 4.455482  \\ \hline

~~~0.09~~~~~~~~~~~~~~~ & 3.893662~~~~~~~~~~~~~~~ & 6.114945~~~~~~~~~~~~~~~ & 3.678032~~~~~~~~~~~~~~~ & 4.535687  \\ \hline
\hline
\end{tabular}
\end{table}

Because the conditions~\eqref{Eq:condition1} and~\eqref{Eq:condition2} cannot be solved analytically in closed form, we determine the MBO and ISCO numerically. In particular, for a fixed charge parameter $Q$, $\omega_{q}$ and varying values of the quintessence parameter $c_q$, we compute the corresponding radii $r_{\rm MBO}$ and $r_{\rm ISCO}$, together with the associated angular momenta $\mathit{L}_{\rm MBO}$ and $\mathit{L}_{\rm ISCO}$. For each parameter choice, the inequality $\mathit{L}_{\rm ISCO}<\mathit{L}_{\rm MBO}$ is satisfied, thereby defining the angular momentum window within which bound periodic trajectories can exist~\cite{Levin:2008mq,Azreg-Ainou:2020bfl}. The resulting numerical values are listed in Table~\ref{tab:ISCO_IBCO_lambda}.

\begin{figure}[ht!]
\centering
\includegraphics[width=7cm]{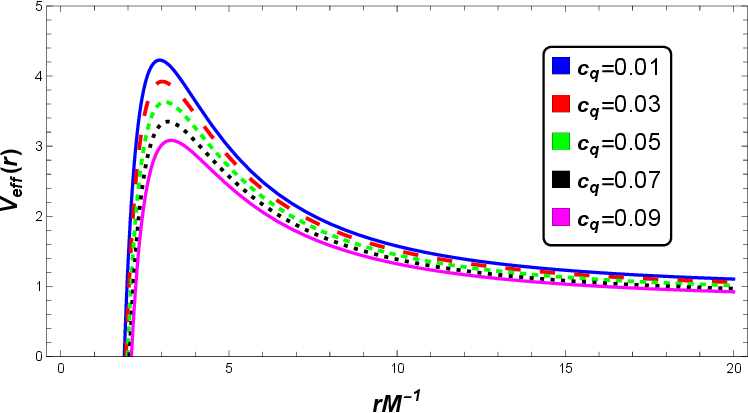}\hspace{1cm}
\includegraphics[width=7cm]{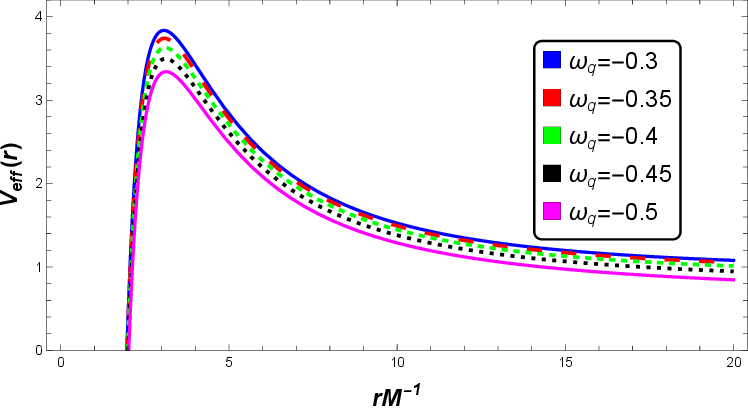}
\includegraphics[width=7cm]{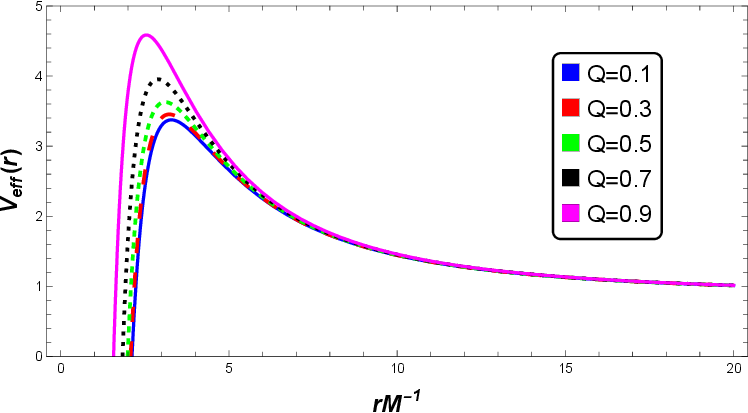}\hspace{1cm}
\includegraphics[width=7cm]{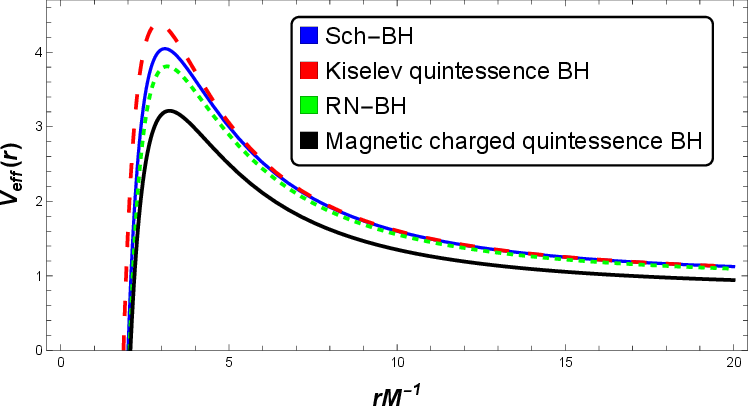}
\caption{Radial profiles of the timelike effective potential $V_{\rm eff}(r)$ for different values of the magnetically charged immersed with quintessence BH parameters. In the top-left panel, we fix $Q=0.5$ and $\omega_q=-0.4$ and vary the quintessence strength parameter $c_q$. In the top-right panel, we fix $Q=0.5$ and $c_q=0.05$ and vary the equation of state parameter $\omega_q$. In the bottom-left panel, we fix $\omega_q=-0.4$ and $c_q=0.05$ and vary the magnetic charge parameter $Q$. The bottom-right panel compares the effective potential of the present magnetically charged BH in a quintessence background with the Schwarzschild, RN, and Kiselev BH geometries.}\label{fig:veff}
\end{figure}
Figure~\ref{fig:veff} displays the radial profile of the timelike effective potential for different choices of the BH parameters. The top-left panel shows that increasing the quintessence strength parameter $c_q$ lowers the peak of the effective potential and weakens the potential barrier. A similar behavior is observed in the top-right panel: as the equation of state parameter $\omega_q$ takes more negative values, the effective potential is further reduced. In contrast, the bottom-left panel shows that increasing the magnetic charge parameter $Q$ enhances the height of the potential barrier, thereby strengthening the radial confinement of timelike motion.

These trends have direct implications for the structure of bound timelike orbits. Since the extrema of $V_{\rm eff}$ determine the location and stability of circular orbits, changes in $c_q$, $\omega_q$, and $Q$ modify the allowed domain of bound trajectories and, consequently, the structure of periodic motion.

\begin{figure}[ht!]
\centering
\includegraphics[width=7cm]{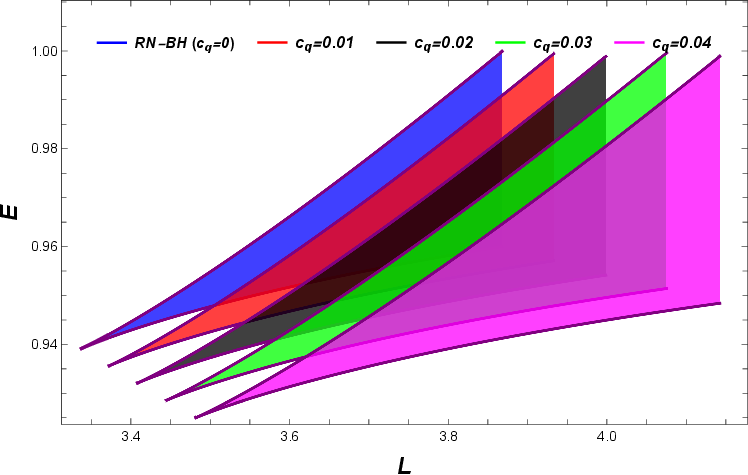}\hspace{1cm}
\includegraphics[width=7cm]{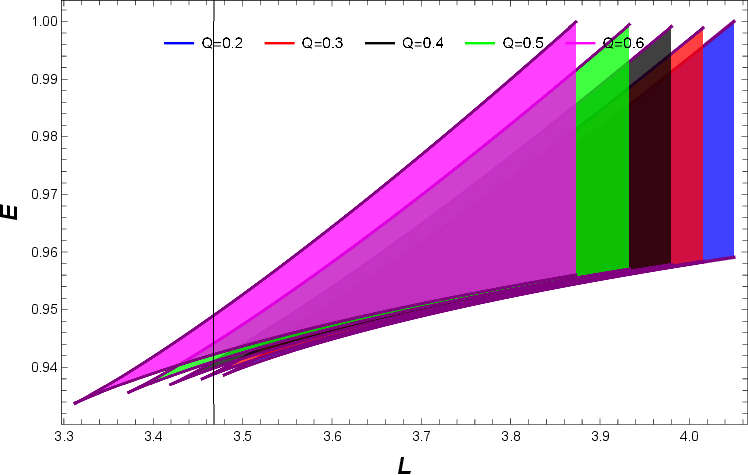}
\caption{Allowed $\mathit{L}-\mathit{E}$ parameter space for bound timelike orbits around the magnetically charged BH in a quintessence background. In the left panel, we fix $Q=0.5$ and $\omega_q=-0.5$ and vary the quintessence strength parameter $c_q$. In the right panel, we fix $c_q=0.02$ and $\omega_q=-0.5$ and vary the magnetic charge parameter $Q$.}\label{fig:L-E}
\end{figure}

The bottom-right panel provides a comparative picture of the effective potential for several benchmark spacetimes. For the parameter choice shown, the magnetically charged BH in a quintessence background exhibits the lowest effective potential profile among the geometries considered, while the Kiselev BH has the highest peak. This comparison highlights the nontrivial interplay between magnetic charge and quintessential matter: their combined effect deforms the strong-field potential more significantly than either the Schwarzschild, RN, or pure Kiselev backgrounds taken separately. Since the effective potential controls the existence of circular and periodic timelike orbits, these differences are expected to be reflected in the corresponding orbital taxonomy and in the GW signatures emitted by particles moving in such backgrounds.
\begin{figure}[ht!]
\centering
\includegraphics[width=7cm]{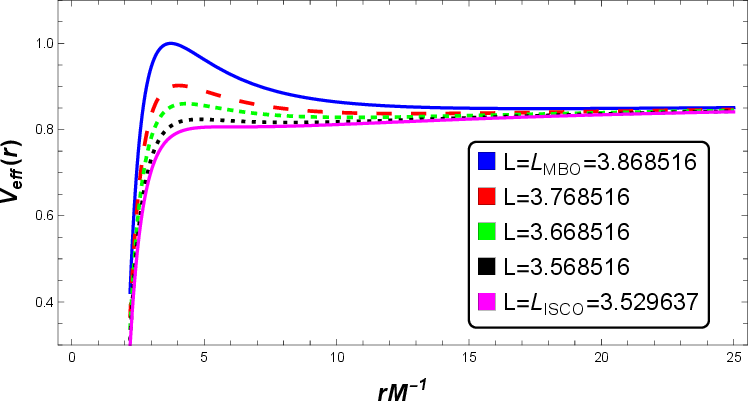}\hspace{1cm}
\caption{Radial profile of the timelike effective potential for several values of the orbital angular momentum $\mathit{L}$, with the BH parameters fixed at $Q=0.5$, $c_q=0.05$, and $\omega_q=-0.4$. The blue curve corresponds to the marginally bound orbit $\mathit{L}_{\rm MBO}=3.868516$ for which the effective potential exhibits two extrema. As $\mathit{L}$ decreases, the local maximum and minimum move toward one another, and they merge at the ISCO $\mathit{L}_{\rm ISCO}=3.529637$ (magenta curve), marking the transition to marginal stability.}\label{fig:veff-1}
\end{figure}
For  fixed BH parameters, bound timelike motion is possible only when the particle angular momentum lies in the interval $\mathit{L}_{\rm ISCO}\leq \mathit{L}\leq\mathit{L}_{\rm MBO}$.
This behavior is illustrated in Fig.~\ref{fig:veff-1}, which shows the effective potential for several values of $\mathit{L}$. At $\mathit{L}=\mathit{L}_{\rm MBO}=3.868516$, the potential possesses two extrema, corresponding to the marginally bound configuration. As the angular momentum decreases, these extrema move progressively closer to each other, and they eventually merge at $\mathit{L}=\mathit{L}_{\rm ISCO}=3.529637$, where the circular orbit becomes marginally stable. Consequently, the interval between $\mathit{L}_{\rm ISCO}$ and $\mathit{L}_{\rm MBO}$ defines the range of angular momentum for which a potential well exists. In addition, the particle energy lies within this well, the motion is bound and can support periodic trajectories.
Bound orbits exist only when the angular momentum lies in the allowed interval and the particle energy remains within the corresponding potential well. The BH parameters therefore determine the allowed region in the $\mathit{L}-\mathit{E}$ plane for bound timelike motion. In Fig.~\ref{fig:L-E}, we display this admissible parameter space for different values of the quintessence strength parameter $c_q$ and the magnetic charge $Q$.

The left panel of Fig.~\ref{fig:L-E} shows that increasing $c_q$ shifts the allowed bound orbit region toward larger values of the angular momentum. A similar trend is seen in the right panel, where the admissible $\mathit{L}-\mathit{E}$ domain also moves to higher $\mathit{L}$ as the magnetic charge $Q$ increases. Thus, for fixed particle energy, both the quintessence background and the magnetic charge increase the upper angular momentum range compatible with bound timelike motion.
These changes in the $\mathit{L}-\mathit{E}$ parameter space are directly relevant to the existence and classification of periodic trajectories, to which we turn in the next section.

\section{Effects of magnetic charge and quintessence on periodic orbits}\label{sec:Periodic Orbits}
We now turn to a distinguished subclass of bound timelike trajectories, namely periodic orbits in the magnetically charged BH in a quintessence background spacetime. For equatorial motion in a static and spherically symmetric background, the orbital dynamics is completely specified by the radial and azimuthal coordinates $(r,~\phi)$. A bound trajectory oscillates between two turning points, (say) $r_1$ and $r_2$, and the corresponding apsidal angle accumulated during one radial cycle is given by~\cite{Li:2024tld,misner1973gravitation,Qi:2024tim,chandrasekhar1998mathematical}
\begin{equation}
\Delta\phi = \oint d\phi = 2 \int_{r_1}^{r_2} \frac{d\phi}{dr} dr. \label{Eq:integration}
\end{equation}
Periodic orbits are closed, bound trajectories for which the radial and azimuthal motions are commensurate so the ratio of the corresponding frequencies is rational. To classify such orbits, we adopt the Levin-Perez-Giz taxonomy~\cite{Levin:2008mq}, in which each periodic trajectory is characterized by a triplet of integers $(\mathit{z},\mathit{w},\mathit{v})$ describing its zoom, whirl, and vertex structure. This framework has been applied extensively to BH spacetimes, including the Schwarzschild and Kerr BHs along with other geometries~\cite{Levin:2009sk,Misra:2010pu,Babar:2017gsg, Tu:2023xab,Li:2024tld,Deng:2020yfm,Lin:2021noq,Gao:2021arw,Zhang:2022psr,Lin:2022wda,Habibina:2022ztd,Wang:2022tfo,Yao:2023ziq,Lin:2023eyd,
Chan:2025ocy,Haroon:2025rzx,Wang:2025wob,Alloqulov:2025bxh,Wei:2025qlh}, and provides a natural basis for analyzing the orbital structure and the associated GW emission from periodic motion.

According to the Levin-Perez-Giz taxonomy~\cite{Levin:2008mq}, each bound periodic orbit can be characterized by a rational number $q_{\rm orb}$, defined as
\begin{equation}
q_{\rm orb} = \frac{\Delta\phi}{2\pi} - 1 = \mathit{w} + \frac{\mathit{v}}{\mathit{z}}. \label{Eq:q-value}
\end{equation}

The integers $(\mathit{z},\mathit{w},\mathit{v})$ admit a simple geometric interpretation: $\mathit{z}$ counts the number of zooms, $\mathit{w}$ the number of whirls near the center, and $\mathit{v}$ specifies the order in which the vertices are traced. To avoid degeneracy, $\mathit{z}$ and $\mathit{v}$ must be relatively prime~\cite{Levin:2008mq}. The corresponding rational number $q_{\rm orb}$ is defined by
\begin{equation}
q_{\rm orb}= \frac{\Delta\phi}{2\pi} - 1 = \frac{1}{\pi} \int_{r_1}^{r_2} \frac{d\phi}{dr} \, dr=\frac{1}{\pi} \int_{r_1}^{r_2} \frac{\mathit{L}}{r^2 \sqrt{\mathit{E}^2 - f(r)\left(1 + \frac{\mathit{L}^2}{r^2}\right)}} \, dr - 1.
\end{equation}

\begin{figure}[ht!]
\centering
\includegraphics[width=7cm]{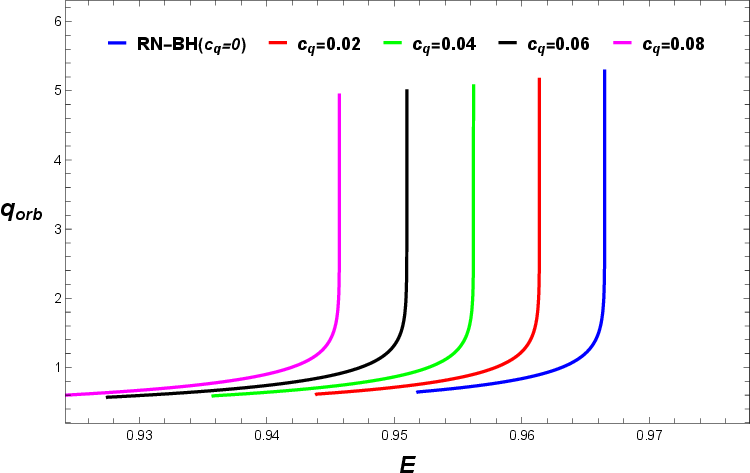}\hspace{1cm}
\includegraphics[width=7cm]{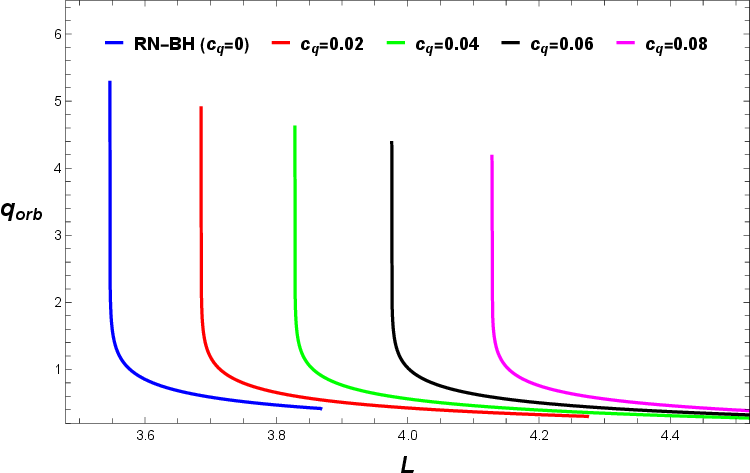}\hspace{1cm}
\caption{Dependence of the rational number $q_{\rm orb}$ on the particle energy $\mathit{E}$ and orbital angular momentum $\mathit{L}$ for different values of the quintessence strength parameter $c_q$. In the left panel, we fix $Q=0.5$, $\omega_q=-0.35$, and $\mathit{L}=\mathit{L}_{\rm MBO}+\mathit{L}_{\rm ISCO})/2$, and show $q_{\rm orb}$ as a function of $\mathit{E}$. In the right panel, we fix $Q=0.5$, $\omega_q=-0.35$, and $E=0.96$, and display $q_{\rm orb}$ as a function of $\mathit{E}$.}\label{fig:q-orb-parameter}
\end{figure}
Figure~\ref{fig:L-E} shows the dependence of the rational number $q_{\rm orb}$ on the particle energy $\mathit{E}$ and orbital angular momentum $\mathit{L}$ for different values of the quintessence parameter $c_q$. In the left panel, $q_{\rm orb}$ increases monotonically with $\mathit{E}$ and grows rapidly as the energy approaches its upper allowed value. At the same time, the curves shift toward lower energies as $c_q$ increases. The right panel displays $q_{\rm orb}$ as a function of $\mathit{L}$. In this case, $q_{\rm orb}$ decreases with increasing angular momentum, while larger values of $c_q$ shift the curves toward higher $L$. These trends show that the quintessence background modifies the periodic orbit spectrum by changing the energy and angular momentum scales at which a given rational orbit can occur.
\begin{table}[htbp]
\centering
\renewcommand{\arraystretch}{1.2}
\begin{tabular}{|c c c c c c c| }
\hline \hline
~~~$c_{q}$~~~~~~~~~~~~~ & $\mathit{L}$~~~~~~~~~~~~~ & $\mathit{E}_{(1,1,0)}$~~~~~~~~~~~~~ & $\mathit{E}_{(1,2,0)}$~~~~~~~~~~~~~ & $\mathit{E}_{(2,1,1)}$~~~~~~~~~~~~~ & $\mathit{E}_{(3,1,2)}$~~~~~~~~~~~~~ & $\mathit{E}_{(4,1,3)}$ \\ [0.5ex] \hline

~~~0   ~~~~~~~~~~~~~ & 3.603126~~~~~~~~~~~~~ & 0.963140~~~~~~~~~~~~~ & 0.966434~~~~~~~~~~~~~ & 0.966026~~~~~~~~~~~~~ & 0.966255~~~~~~~~~~~~~ & 0.966320 \\ [0.5ex] \hline

~~~0.02~~~~~~~~~~~~~ & 3.716704~~~~~~~~~~~~~ & 0.957960~~~~~~~~~~~~~ & 0.961321~~~~~~~~~~~~~ & 0.960920~~~~~~~~~~~~~ & 0.961144~~~~~~~~~~~~~ & 0.961209 \\ [0.5ex] \hline

~~~0.04~~~~~~~~~~~~~ & 3.835287~~~~~~~~~~~~~ & 0.952689~~~~~~~~~~~~~ & 0.956150~~~~~~~~~~~~~ & 0.955741~~~~~~~~~~~~~ & 0.955973~~~~~~~~~~~~~ & 0.956039 \\ [0.5ex] \hline

~~~0.06~~~~~~~~~~~~~ & 3.959304~~~~~~~~~~~~~ & 0.947349~~~~~~~~~~~~~ & 0.950921~~~~~~~~~~~~~ & 0.950507~~~~~~~~~~~~~ & 0.950742~~~~~~~~~~~~~ & 0.950809 \\ [0.5ex] \hline
\hline
\end{tabular}
\caption{Numerically computed energies $\mathit{E}_{(\mathit{z},\mathit{w},\mathit{v})}$ for selected periodic orbits at fixed angular momentum $\mathit{L}=(\mathit{L}_{\rm MBO}+\mathit{L}_{\rm ISCO})/2$, for different values of the quintessence parameter $c_q$.}
\label{tab:energy_levels_cq}
\end{table}

\begin{table}[htbp]
\centering
\renewcommand{\arraystretch}{1.2}
\begin{tabular}{|c c c c c c c| }
\hline \hline
~~~$c_{q}$~~~~~~~~~~~~~ & $\mathit{E}$~~~~~~~~~~~~~ & $\mathit{L}_{(1,1,0)}$~~~~~~~~~~~~~ & $\mathit{L}_{(1,2,0)}$~~~~~~~~~~~~~ & $\mathit{L}_{(2,1,1)}$~~~~~~~~~~~~~ & $\mathit{L}_{(3,1,2)}$~~~~~~~~~~~~~ & $\mathit{L}_{(4,1,3)}$ \\ [0.5ex] \hline

~~~0   ~~~~~~~~~~~~~ & 0.96~~~~~~~~~~~~~ & 3.577101~~~~~~~~~~~~~ & 3.547399~~~~~~~~~~~~~ & 3.551499~~~~~~~~~~~~~ & 3.549277~~~~~~~~~~~~~ & 3.548599 \\ [0.5ex] \hline

~~~0.02~~~~~~~~~~~~~ & 0.96~~~~~~~~~~~~~ & 3.714110~~~~~~~~~~~~~ & 3.686155~~~~~~~~~~~~~ & 3.689699~~~~~~~~~~~~~ & 3.687755~~~~~~~~~~~~~ & 3.687170 \\ [0.5ex] \hline

~~~0.04~~~~~~~~~~~~~ & 0.96~~~~~~~~~~~~~ & 3.856131~~~~~~~~~~~~~ & 3.828869~~~~~~~~~~~~~ & 3.832179~~~~~~~~~~~~~ & 3.830307~~~~~~~~~~~~~ & 3.829775 \\ [0.5ex] \hline

~~~0.06~~~~~~~~~~~~~ & 0.96~~~~~~~~~~~~~ & 4.002641~~~~~~~~~~~~~ & 3.976239~~~~~~~~~~~~~ & 3.979299~~~~~~~~~~~~~ & 3.977560~~~~~~~~~~~~~ & 3.977068 \\ [0.5ex] \hline
\hline
\end{tabular}
\caption{Numerically computed angular momenta $\mathit{L}_{(\mathit{z},\mathit{w},\mathit{v})}$ for selected periodic orbits at fixed energy $E=0.96$, for different values of the quintessence parameter $c_q$.}
\label{tab:angular_levels_cq}
\end{table}
To characterize the periodic trajectories in the present spacetime, we numerically determine the energies $\mathit{E}_{(\mathit{z},\mathit{w},\mathit{v})}$ and angular momenta $\mathit{L}_{(\mathit{z},\mathit{w},\mathit{v})}$ associated with selected periodic orbits. In particular, the energies $\mathit{E}_{(\mathit{z},\mathit{w},\mathit{v})}$ are computed at fixed angular momentum $\mathit{L}=(\mathit{L}_{\rm MBO}+\mathit{L}_{\rm ISCO})/2$, while the corresponding angular momenta $\mathit{L}_{(\mathit{z},\mathit{w},\mathit{v})}$ are obtained at fixed energy $\mathit{E}=0.96$. The resulting values for several representative periodic orbits are listed in Tables~\ref{tab:energy_levels_cq} and~\ref{tab:angular_levels_cq}, which illustrate how the periodic orbit spectrum shifts with the quintessence parameter $c_q$.

\begin{figure}[ht!]
\centering
\includegraphics[width=4cm]{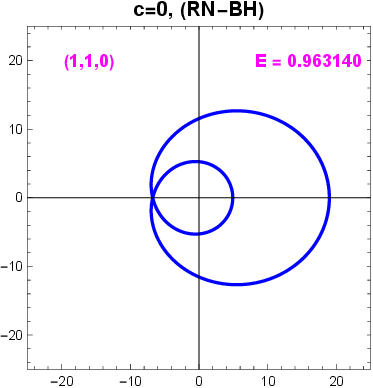}
\includegraphics[width=4cm]{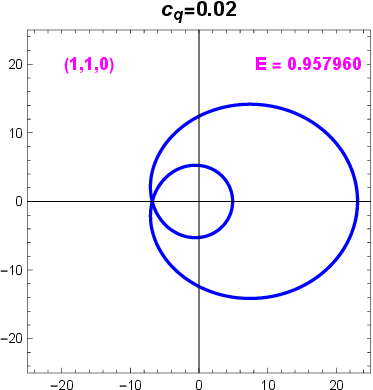}
\includegraphics[width=4cm]{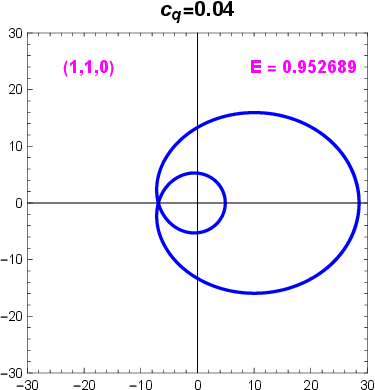}
\includegraphics[width=4cm]{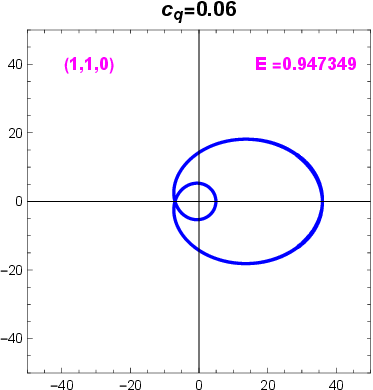}\\
\includegraphics[width=4cm]{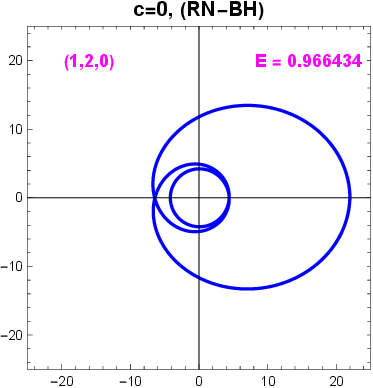}
\includegraphics[width=4cm]{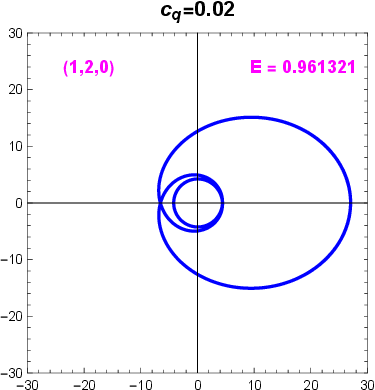}
\includegraphics[width=4cm]{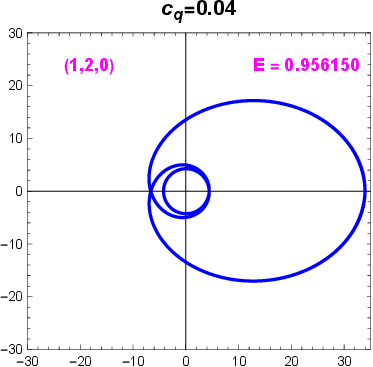}
\includegraphics[width=4cm]{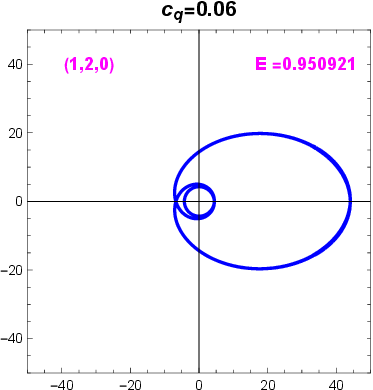}\\
\includegraphics[width=4cm]{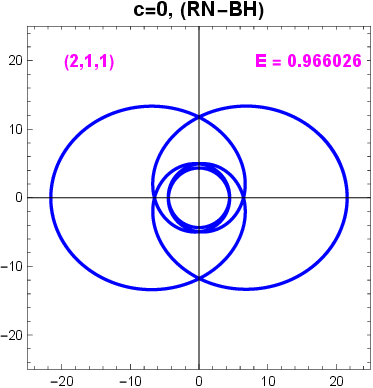}
\includegraphics[width=4cm]{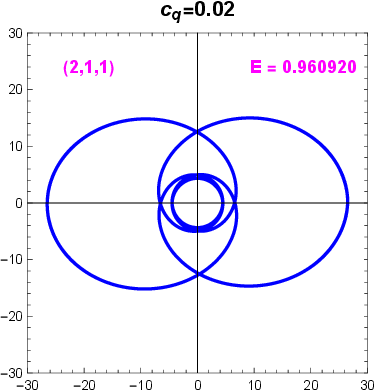}
\includegraphics[width=4cm]{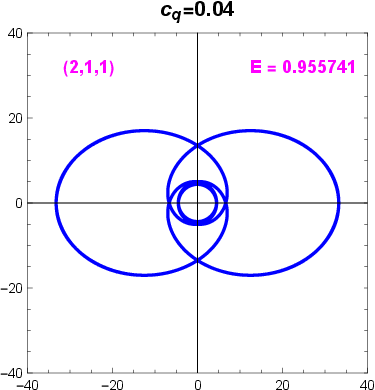}
\includegraphics[width=4cm]{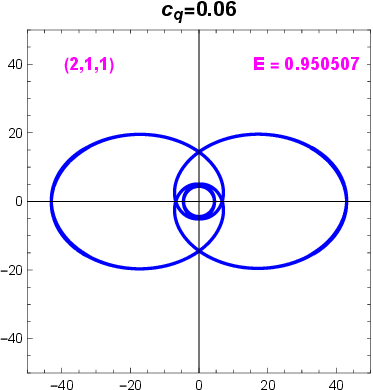}\\
\includegraphics[width=4cm]{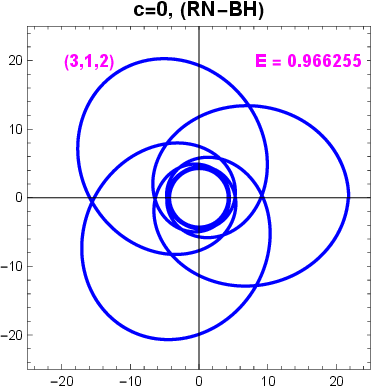}
\includegraphics[width=4cm]{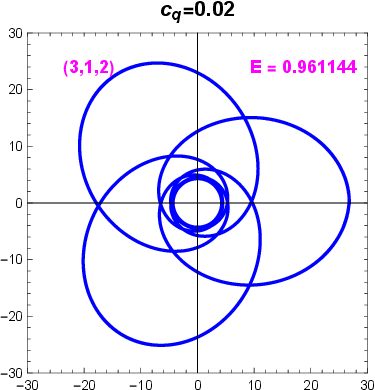}
\includegraphics[width=4cm]{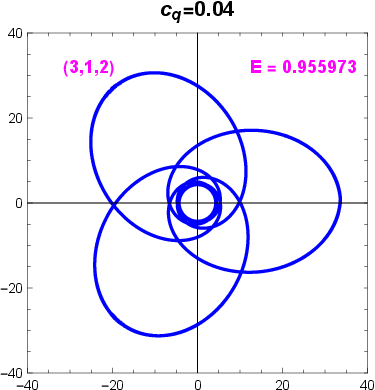}
\includegraphics[width=4cm]{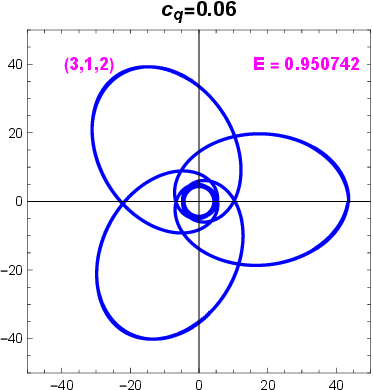}\\
\includegraphics[width=4cm]{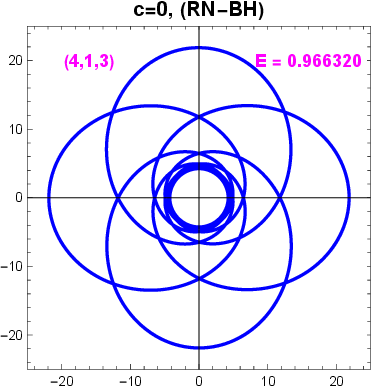}
\includegraphics[width=4cm]{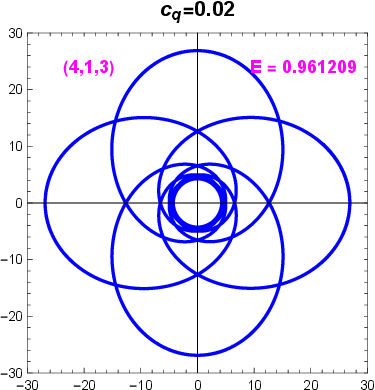}
\includegraphics[width=4cm]{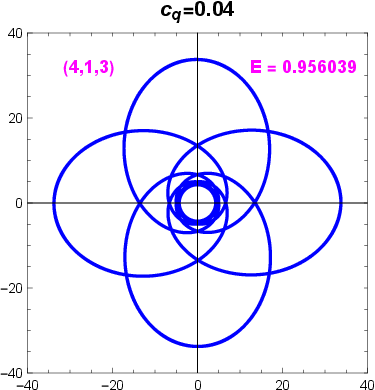}
\includegraphics[width=4cm]{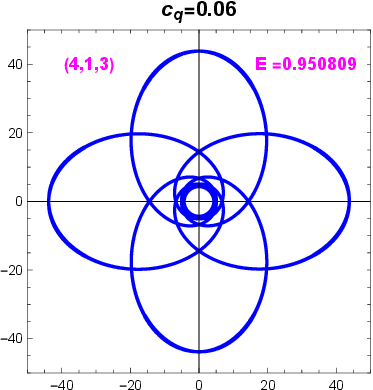}
\caption{Representative periodic timelike orbits obtained from the energies $\mathit{L}_{(\mathit{z},\mathit{w},\mathit{v})}$ listed in Table~\ref{tab:energy_levels_cq}, for fixed $Q=0.5$ and $\mathit{L}=(\mathit{L}_{\rm MBO}+\mathit{L}_{\rm ISCO})/2$}\label{fig:E-orbits}
\end{figure}

\begin{figure}[ht!]
\centering
\includegraphics[width=4cm]{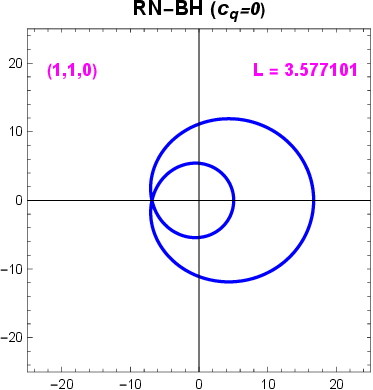}
\includegraphics[width=4cm]{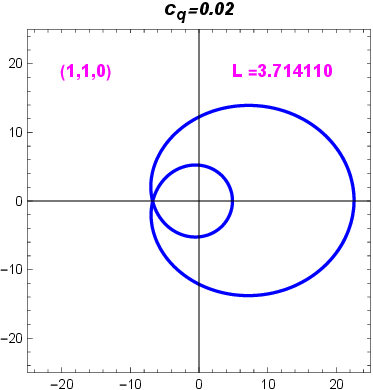}
\includegraphics[width=4cm]{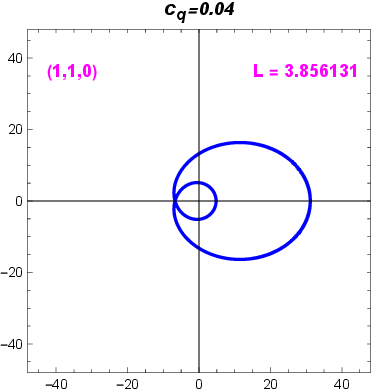}
\includegraphics[width=4cm]{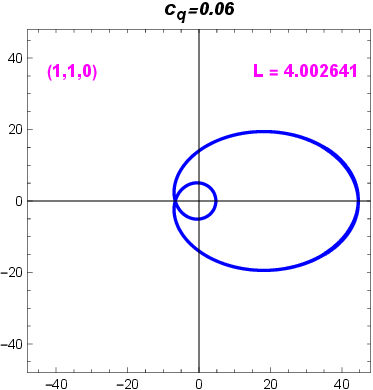}\\
\includegraphics[width=4cm]{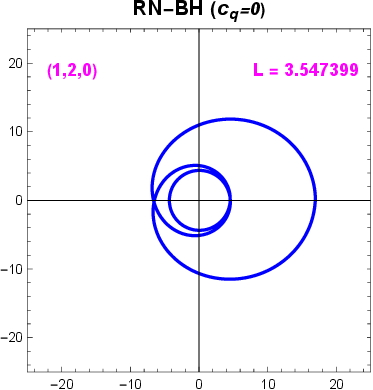}
\includegraphics[width=4cm]{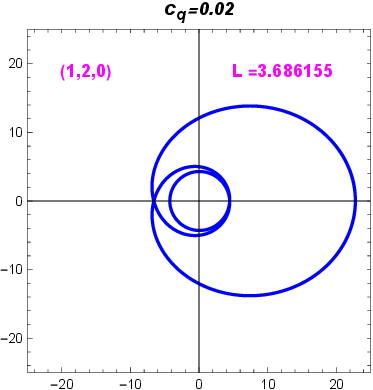}
\includegraphics[width=4cm]{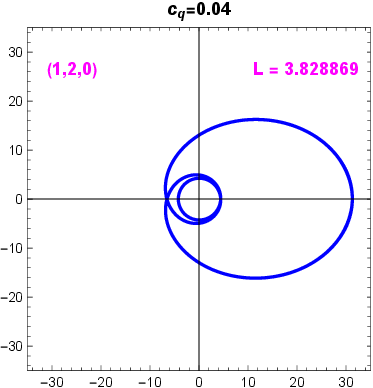}
\includegraphics[width=4cm]{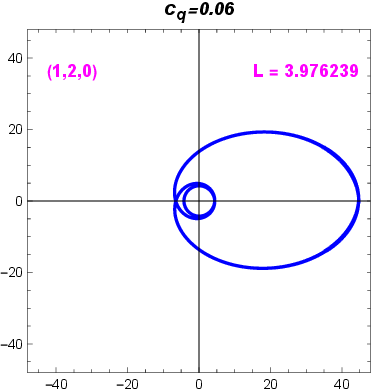}\\
\includegraphics[width=4cm]{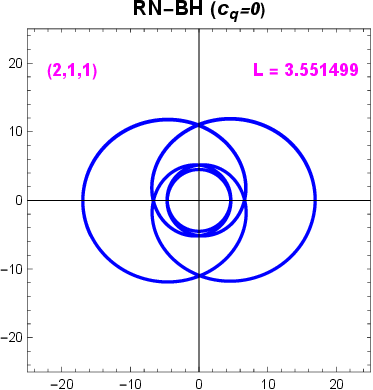}
\includegraphics[width=4cm]{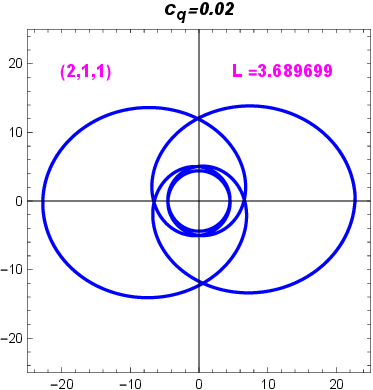}
\includegraphics[width=4cm]{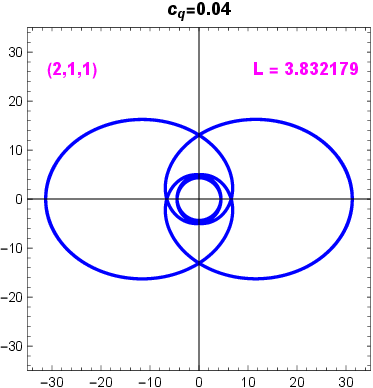}
\includegraphics[width=4cm]{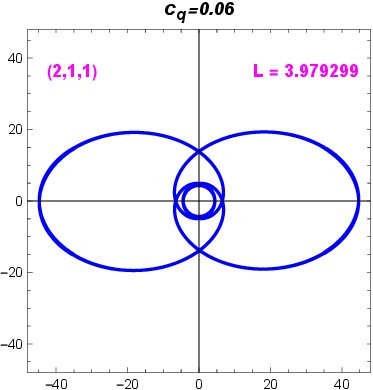}\\
\includegraphics[width=4cm]{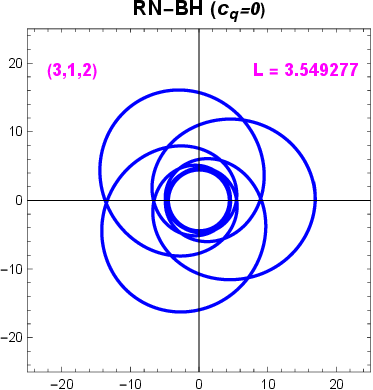}
\includegraphics[width=4cm]{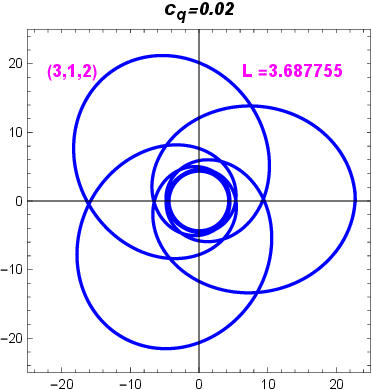}
\includegraphics[width=4cm]{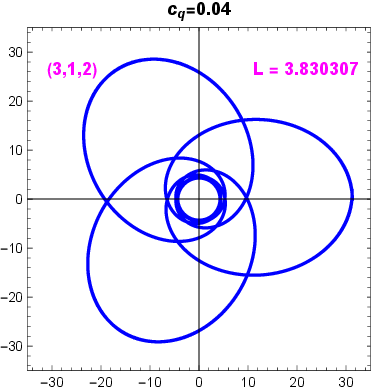}
\includegraphics[width=4cm]{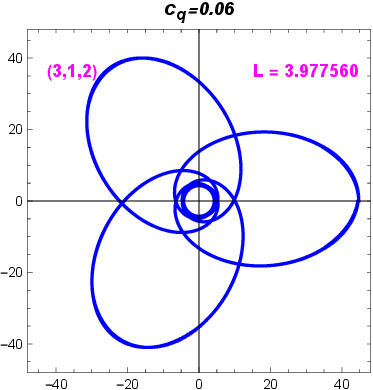}\\
\includegraphics[width=4cm]{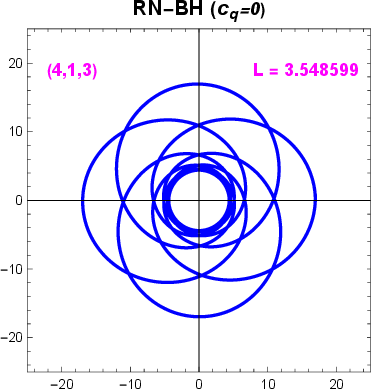}
\includegraphics[width=4cm]{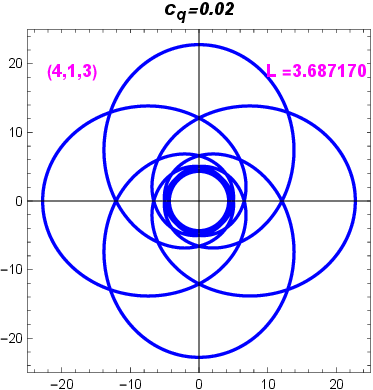}
\includegraphics[width=4cm]{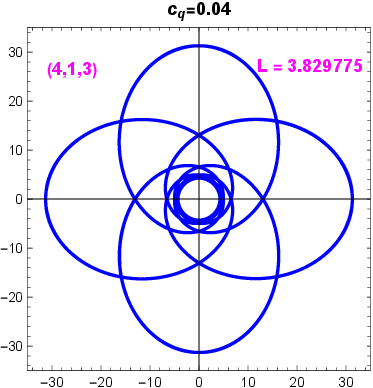}
\includegraphics[width=4cm]{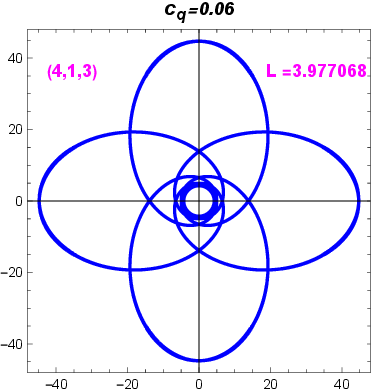}
\caption{Representative periodic timelike orbits obtained from the angular momenta $\mathit{L}_{(\mathit{z},\mathit{w},\mathit{v})}$ listed in Table~\ref{tab:angular_levels_cq}, for fixed $Q=0.5$ and $E=0.96$.}\label{fig:L-orbits}
\end{figure}

Figures~\ref{fig:E-orbits} and~\ref{fig:L-orbits} display the periodic trajectories corresponding to the numerical values listed in Tables~\ref{tab:energy_levels_cq} and~\ref{tab:angular_levels_cq}, respectively. Figure~\ref{fig:E-orbits} shows representative periodic orbits obtained from the energies $\mathit{E}_{(\mathit{z},\mathit{w},\mathit{v})}$ at fixed angular momentum, whereas Fig.~\ref{fig:L-orbits} presents the corresponding trajectories obtained from the angular momenta $\mathit{L}_{(\mathit{z},\mathit{w},\mathit{v})}$ at fixed energy. In both cases, the topological class associated with a given triplet $\mathit{z},\mathit{w},\mathit{v}$ remains unchanged, while the quintessence parameter $c_q$ deforms the geometry of the orbit by modifying its radial extent, turning points, and whirl structure. In particular, increasing $c_q$ systematically changes the orbital scale, showing that the quintessence background alters the strong-field geodesic structure without changing the underlying periodic orbit taxonomy.

Having established the periodic orbit structure, we now investigate the gravitational radiation generated by a test particle moving on such trajectories around a magnetically charged BH in a quintessence background.


%
\section{Numerical kludge gravitational radiation from periodic orbits} \label{sec:GW}
In this section, we study the gravitational radiation generated by a test particle moving on periodic timelike orbits around a magnetically charged BH in a quintessence background. This setup provides an effective description of an EMRI, in which a stellar-mass compact object orbits a supermassive BH. Such systems are among the primary targets of future space-based GW detectors, including LISA, Taiji, and TianQin~\cite{Hu:2017mde,TianQin:2015yph,Gong:2021gvw,Danzmann:1997hm,Schutz:1999xj,Gair:2004iv,LISA:2017pwj,Maselli:2021men}.

We model the small body as a perturbing source moving in the fixed background geometry and adopt the adiabatic approximation, under which the radiation-reaction timescale is much longer than the orbital period~\cite{Hughes:1999bq,Barack:2003fp,Isoyama:2021jjd}. To compute the associated waveforms, we employ the numerical kludge approach~\cite{Babak:2006uv,Tu:2023xab}. In this scheme, the geodesic motion is first obtained numerically in the BH background, after which the GW-form is constructed from the corresponding trajectory using the quadrupole formula. Although approximate, this semirelativistic framework has been shown to be effective for modeling EMRI waveforms and probing the influence of the background geometry on the emitted radiation~\cite{Gair:2004iv,Barack:2003fp,Hughes:2000ssa,Ahmed:2025azu}, the orbit, the properties of the central BH, and possible environmental effects \cite{Gair:2004iv,Barack:2003fp,Hughes:2000ssa,Ahmed:2025azu}.

Within the numerical kludge scheme, the GW-form is constructed from the quadrupole formula. The metric perturbation $\mathit{h}_{ij}$ is related to the second time derivative of the symmetric trace-free mass quadrupole moment $\mathit{I}_{ij}$ through~\cite{Maselli:2021men},
\begin{equation}
\mathit{h}_{ij} = \frac{1}{A}\mathit{\ddot{I}}_{ij}, \label{Eq:GW-1}
\end{equation}
where $A = c^4 D_L/(2G)$, and $D_L$ is the luminosity distance to the source, and we adopt geometrized units $G=c=1$.
The particle trajectory $\mathit{Z}_{i}(t)$ entering the waveform calculation is obtained by numerically solving the geodesic equations in the BH spacetime.
For a particle of mass $m$, the mass quadrupole moment is \cite{Thorne:1980ru}
\begin{equation}
\mathit{I}^{ij} = m \int d^3x \, x^i x^j \, \delta^3(x^i - \mathit{Z}^{i}(t)). \label{Eq:GW-2}
\end{equation}

Writing the motion in Cartesian coordinates,
\begin{equation}
x = r \sin\theta \cos\phi, \quad y = r \sin\theta \sin\phi, \quad z = r \cos\theta. \label{Eq:GW-3}
\end{equation}
The quadrupole waveform takes the form
\begin{equation}
\mathit{h}_{ij} = \frac{m}{A} \left( a_i x_j + a_j x_i + 2 v_i v_j \right), \label{Eq:GW-4}
\end{equation}
where $v_i$ and $a_i$ are the Cartesian velocity and acceleration components of the particle. To extract the observable polarizations, we introduce a detector frame $(X,Y,Z)$ with basis vectors~\cite{Babak:2006uv}
The unit vectors of the detector frame expressed in the source coordinates $(x,y,z)$ are
\begin{align}
\hat{e}_X &= (\cos\xi, -\sin\xi, 0), \\
\hat{e}_Y &= (\sin\iota\sin\xi, \cos\iota\cos\xi, -\sin\iota),\\
\hat{e}_Z &= (\sin\iota\sin\xi, -\sin\iota\cos\xi, \cos\iota),
\end{align} \label{Eq:GW-5}
from which the two GW polarizations are obtained as
\begin{align}
\mathit{h}_+ &= \frac{1}{2}(e_X^i e_X^j - e_Y^i e_Y^j)\mathit{h}_{ij},  \\
\mathit{h}_\times &= \frac{1}{2}(e_X^i e_Y^j - e_Y^i e_X^j)\mathit{h}_{ij}.
\end{align} \label{Eq:GW-6}
These polarizations can be expressed in terms of components $\mathit{h}_{\iota\xi}$, $\mathit{h}_{\xi\xi}$, and $\mathit{h}_{\iota \iota}$, which are defined in the detector frame as combinations of the $\mathit{h}_{ij}$ components as
\begin{align}\nonumber
h_+ &= \frac{1}{2}(h_{\xi\xi} - h_{\iota\iota}), \\
h_\times &= h_{\iota\xi}, \label{Eq:GW-7}
\end{align}
where the corresponding components are \cite{Babak:2006uv}
\begin{align}
\mathit{h}_{\xi\xi} &= \mathit{h}_{xx} \cos^2\xi - \mathit{h}_{xy} \sin 2\xi + \mathit{h}_{yy} \sin^2\xi, \\
\mathit{h}_{\iota\iota} &= \cos^2\iota [\mathit{h}_{xx} \sin^2\xi + \mathit{h}_{xy} \sin 2\xi + h_{yy} \cos^2\xi] \nonumber \\
&\quad +\mathit{h}_{zz} \sin^2\iota - \sin 2\iota [\mathit{h}_{xz} \sin\xi + \mathit{h}_{yz} \cos\xi], \\
\mathit{h}_{\iota\xi} &= \frac{1}{2} \cos\iota [\mathit{h}_{xx} \sin 2\xi + 2\mathit{h}_{xy} \cos 2\xi -\mathit{h}_{yy} \sin 2\xi] \nonumber \\
&\quad + \sin\iota [\mathit{h}_{yz} \sin\xi - \mathit{h}_{xx} \cos\xi].
\end{align} \label{Eq:GW-8}

To examine the effect of the BH parameters on the waveforms generated by periodic EMRI orbits, we consider a compact object of mass $m=10M_\odot$  orbiting a supermassive BH of mass $M=10^{7}M_\odot$. Unless otherwise stated, we set the inclination angle and longitude of pericenter to $\iota=\xi=\pi/4$, and fix the luminosity distance $D_L$ when evaluating the polarizations $h_+$ and $h_\times$.

The resulting waveforms exhibit the characteristic burst-like structure of zoom-whirl motion. During the zoom phases, when the particle moves along the more extended eccentric part of the orbit, the waveform amplitude remains comparatively small. By contrast, the whirl phases near the central BH generate short, high-amplitude bursts. Thus, the number and spacing of the bursts encode the underlying $\mathit{z},\mathit{w},\mathit{v}$ structure of the periodic orbit.


\begin{figure}[ht!]
\centering
\begin{minipage}[c]{0.5\textwidth}
\centering
\includegraphics[width=\linewidth]{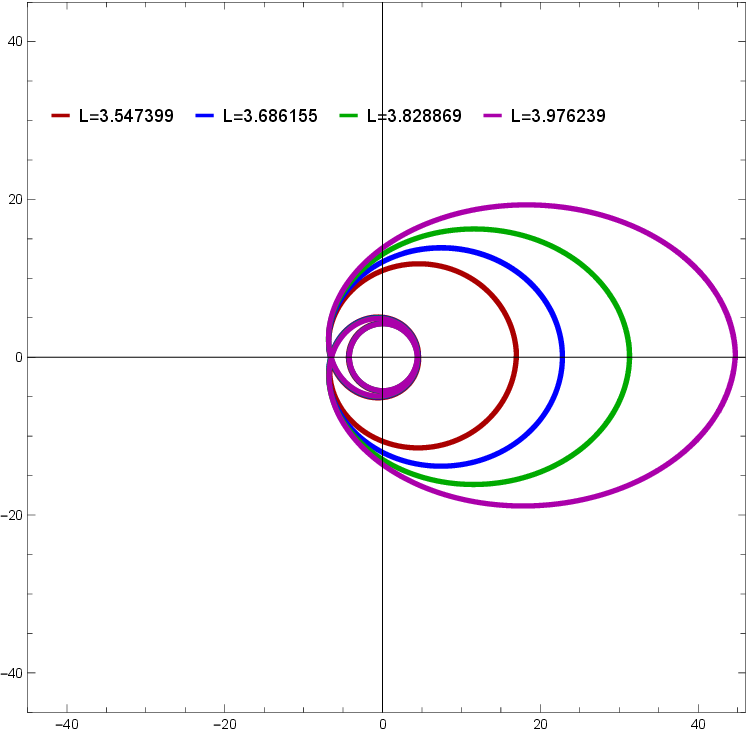}
\end{minipage}%
\hfill
\begin{minipage}[c]{0.5\textwidth}
\centering
\includegraphics[width=\linewidth]{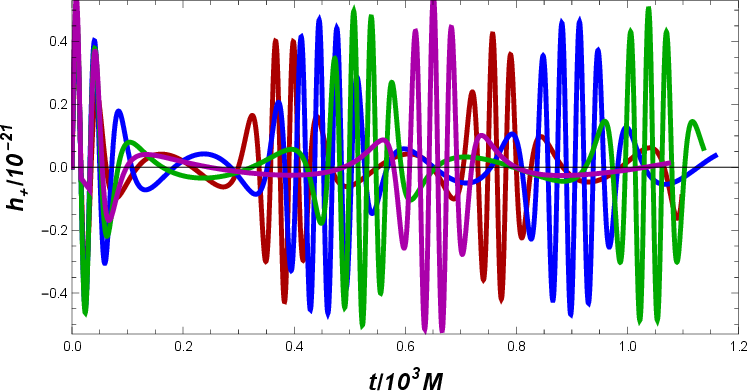}
\includegraphics[width=\linewidth]{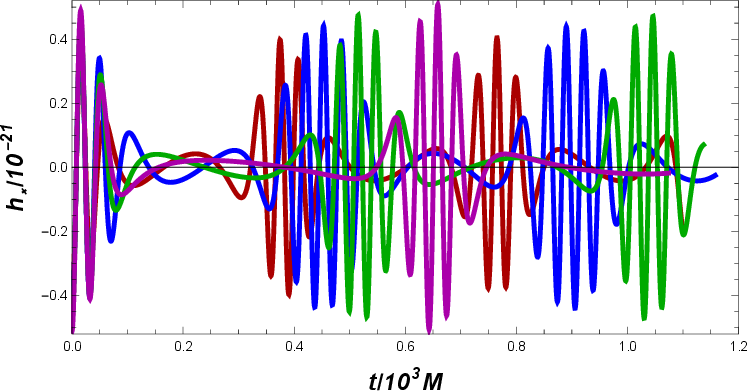}
\end{minipage}
\caption{The $(1,2,0)$ periodic orbit and its associated GW signal for an EMRI system consisting of a stellar-mass compact object orbiting a supermassive magnetically charged BH in a quintessence background. The masses are fixed to $M=10^{7}M_{\odot}$ and $m=10M_{\odot}$, with $Q=0.5$ and $\omega_q=-0.3$. The quintessence parameter is varied as $c_q=0$ (red), $0.02$ (blue), $0.04$ (green), and $0.06$ (magenta)}\label{fig:GW-3}
\end{figure}
\begin{figure}[ht!]
\centering
\begin{minipage}[c]{0.5\textwidth}
\centering
\includegraphics[width=\linewidth]{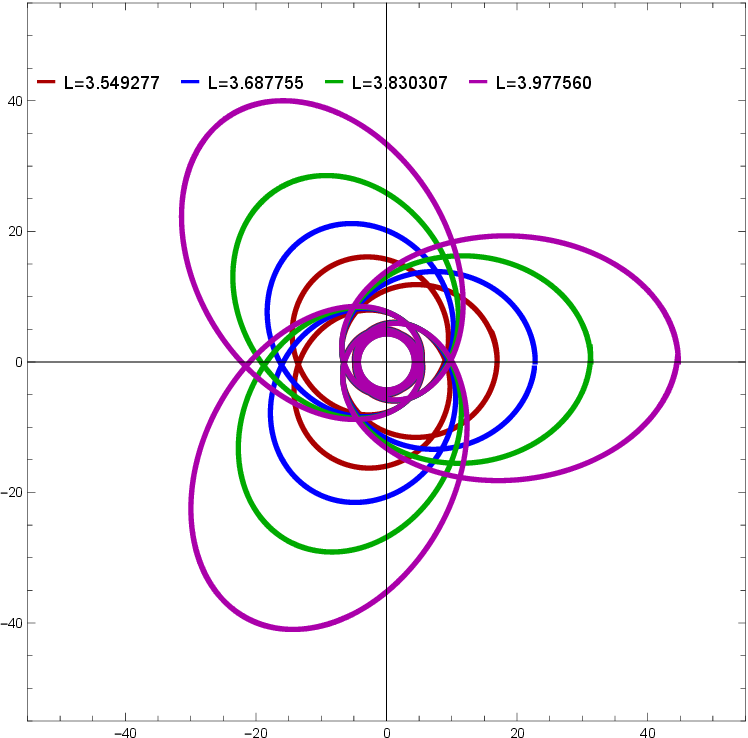}
\end{minipage}%
\hfill
\begin{minipage}[c]{0.5\textwidth}
\centering
\includegraphics[width=\linewidth]{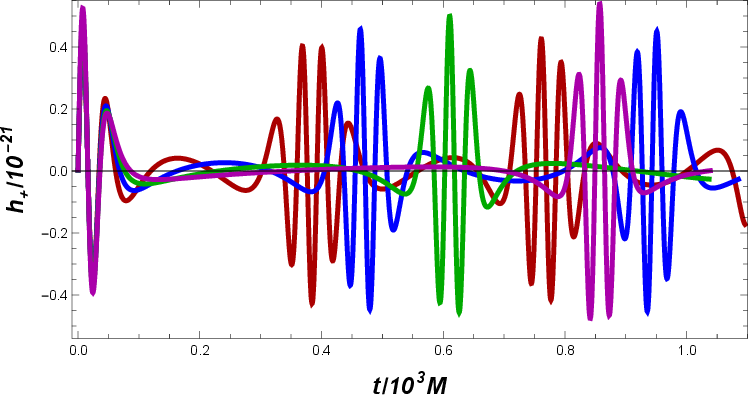}
\includegraphics[width=\linewidth]{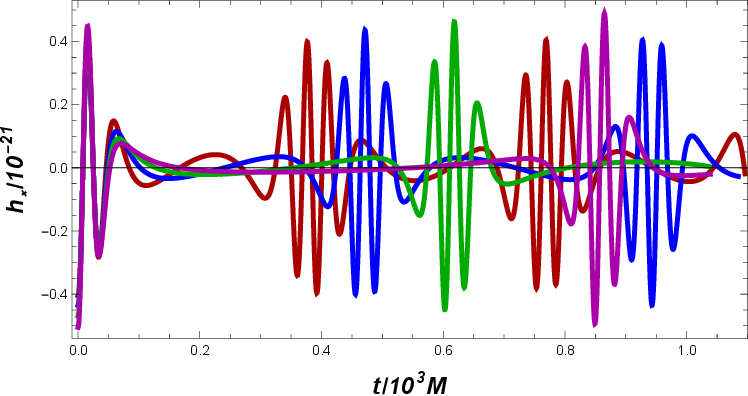}
\end{minipage}
\caption{The plot shows the $(3,1,2)$ periodic orbit and its associated GW signal for the same EMRI configuration as in Fig.~\ref{fig:GW-3}, with identical parameter choices.}\label{fig:GW-4}
\end{figure}

Figures~\ref{fig:GW-3} and~\ref{fig:GW-4} display the waveforms generated by the $(1,2,0)$ and $(3,1,2)$ periodic orbit families, respectively. In each figure, the left panel shows the orbital trajectory, while the right panels present the corresponding polarizations $\mathit{h}_{+}$ and $\mathit{h}_{\times}$ for several values of the quintessence parameter $c_q$. In both families, the waveform exhibits the characteristic zoom-whirl pattern: the low-amplitude, slowly varying part of the signal is produced during the zoom phase, whereas the whirl motion near periapsis generates short bursts with larger amplitude and higher frequency. The more intricate $(3,1,2)$ orbit produces a correspondingly richer burst structure than the simpler $(1,2,0)$ orbit, reflecting the direct imprint of orbital topology on the emitted radiation. Varying $c_q$ preserves the overall waveform morphology for a given topological class but shifts the timing and frequency content of the bursts through its effect on the underlying geodesic motion.

\begin{figure}[ht!]
\centering
\includegraphics[width=7cm]{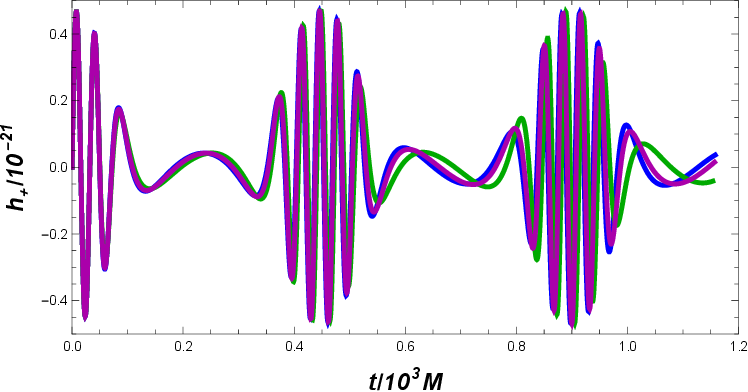}\hspace{1cm}
\includegraphics[width=7cm]{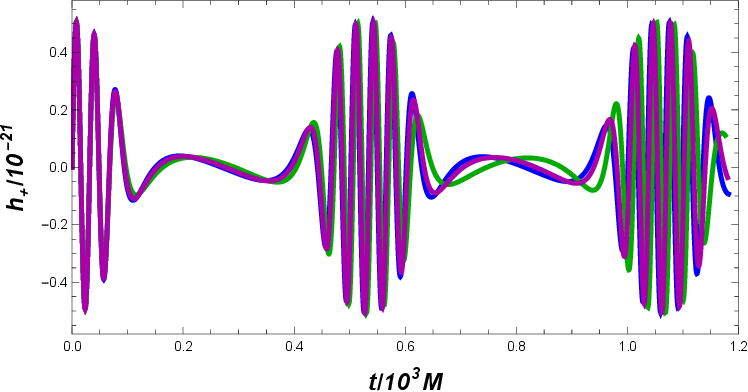}
\includegraphics[width=7cm]{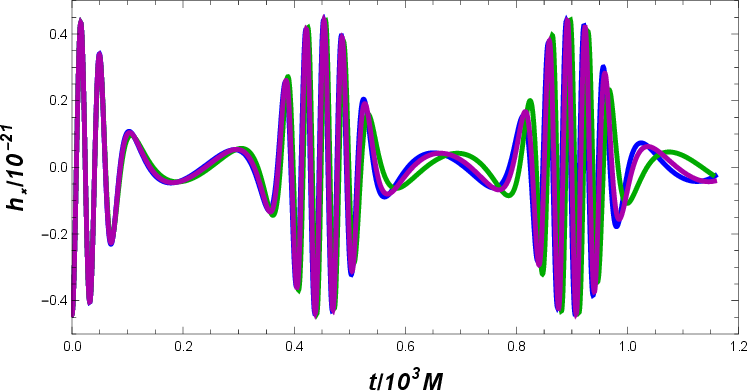}\hspace{1cm}
\includegraphics[width=7cm]{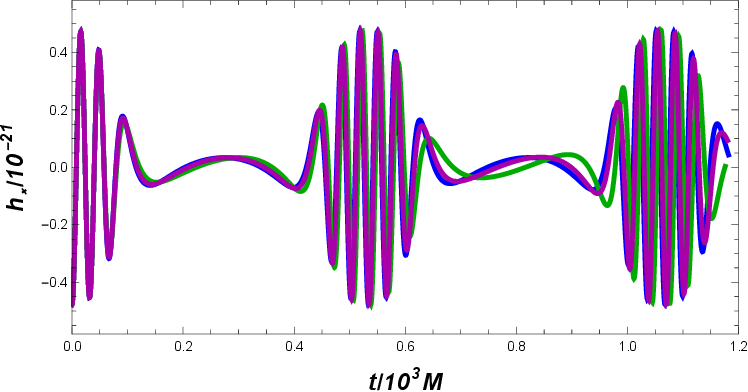}
\caption{Gravitational wave polarizations $\mathit{h}_{+}$ and $\mathit{h}_{\times}$for the periodic-orbit families $(1,2,0)$, $(2,1,1)$, and $(3,1,2)$ in an EMRI system with $M=10^7M_\odot$, $m=10M_\odot$, $Q=0.5$, and $E=0.96$. The upper and lower rows correspond to $c_q=0.02$ and $c_q=0.04$, respectively. The modulation of the signal reflects the underlying zoom-whirl structure of the orbital motion.}\label{fig:GW-1}
\end{figure}

Figure~\ref{fig:GW-1} summarizes the GW-forms generated by three representative periodic orbit families, $(1,2,0)$, $(2,1,1)$, and $(3,1,2)$, for a test particle in an EMRI configuration. The waveforms are computed from Eqs.~\eqref{Eq:GW-4} and~\eqref{Eq:GW-7} and display both polarizations, $\mathit{h}_{+}$ and $\mathit{h}_{\times}$, over one radial period. In each case, the signal exhibits the characteristic zoom-whirl modulation: low-amplitude segments are associated with the zoom phase, while the whirl phase near periapsis produces short bursts of higher amplitude and frequency. The figure therefore makes explicit the close correspondence between the waveform morphology and the underlying periodic-orbit topology.

A comparison between the upper and lower panels shows that varying the quintessence parameter $c_q$ changes both the phase and the detailed burst structure of the signal, while preserving the qualitative waveform pattern associated with a given $(\mathit{z},\mathit{w},\mathit{v})$ orbit. These shifts reflect the sensitivity of the emitted radiation to the background geometry through its effect on the geodesic motion of the small compact object.
The gravitational radiation generated by periodic motion can be further characterized in the frequency domain through the Fourier amplitudes $|\tilde{\mathit{h}}_{+,\times}(f)|$ and the characteristic strain $\mathit{h}_{c}(f)$, defined by~\cite{Babak:2006uv,Hughes:2000ssa,Gair:2004iv,Ahmed:2025azu}
\begin{equation}
\mathit{h}_{c}(f) = \sqrt{2f \left( |\tilde{\mathit{h}}_{+}(f)|^2 + |\tilde{\mathit{h}}_{\times}(f)|^2 \right)}. \label{Eq:GW-9}
\end{equation}

\begin{figure}[ht!]
\centering
\includegraphics[width=7cm]{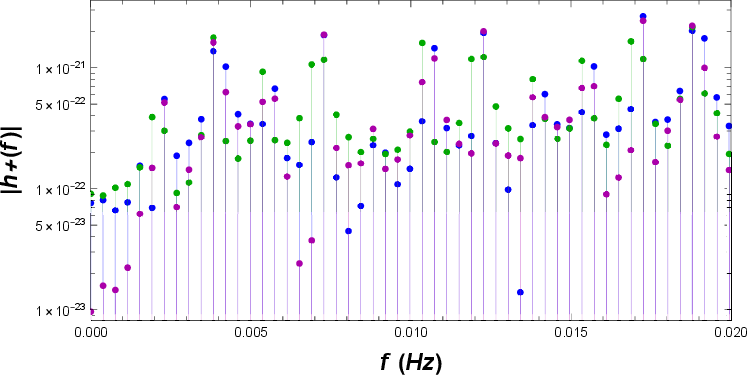}\hspace{1cm}
\includegraphics[width=7cm]{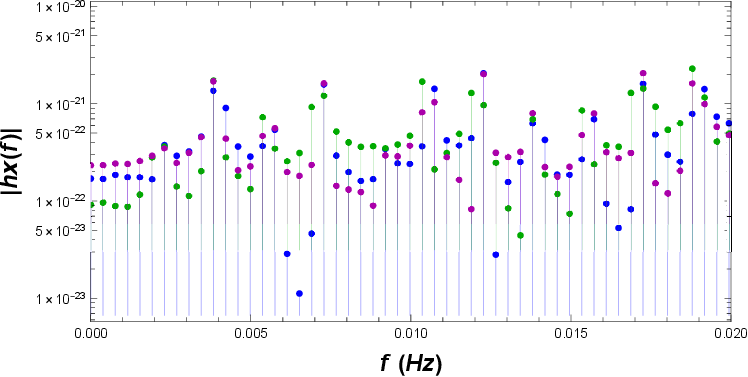}
\includegraphics[width=7cm]{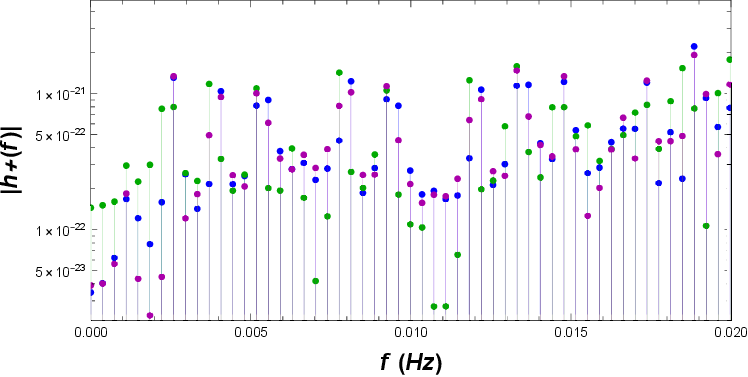}\hspace{1cm}
\includegraphics[width=7cm]{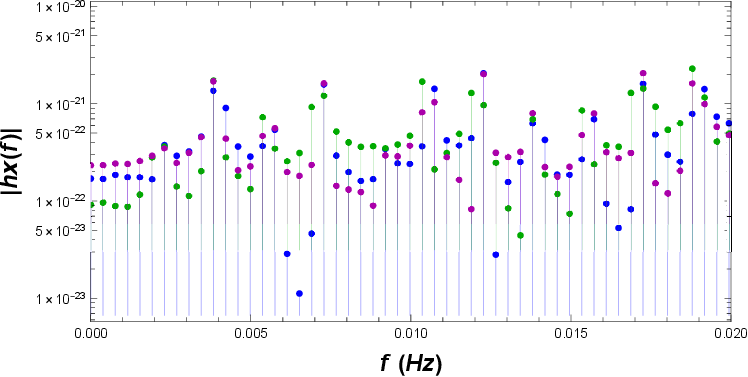}
\caption{Fourier amplitudes $|\tilde{\mathit{h}}_{+,\times}(f)|$ corresponding to the time-domain waveforms of periodic EMRI orbits. The spectra display a discrete harmonic structure associated with the underlying zoom-whirl motion. Variations in the quintessence parameter $c_q$ redistribute spectral power among the harmonic modes and modify the high-frequency content of the signal.}\label{fig:GW-2}
\end{figure}

\begin{figure}[ht!]
\centering
\includegraphics[width=7cm]{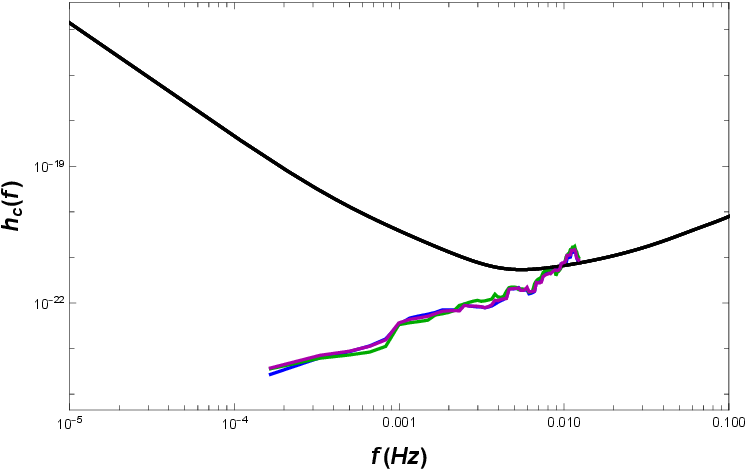}\hspace{1cm}
\includegraphics[width=7cm]{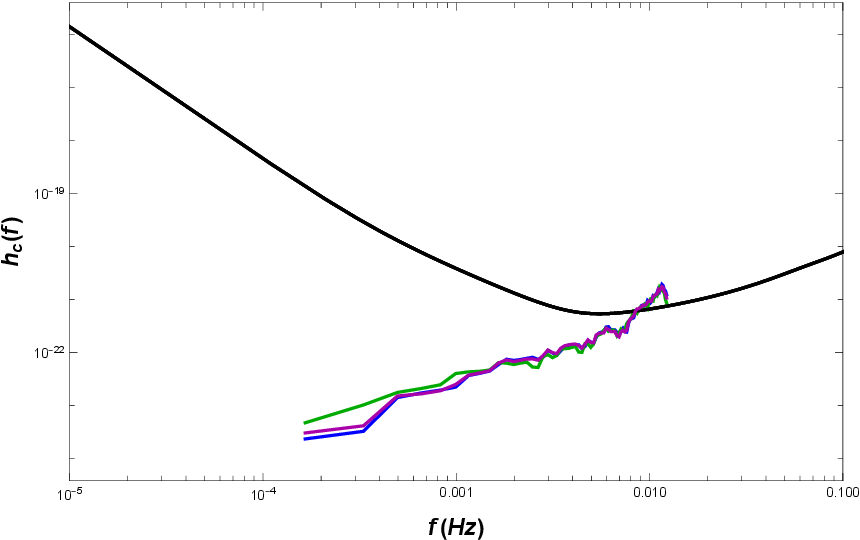}
\caption{Characteristic strain for GWs generated by periodic EMRI orbits in the spacetime of a magnetically charged BH surrounded by quintessence. The black curve denotes the LISA sensitivity. The colored curves show that the strongest harmonic components of the signal lie in the millihertz band and, for part of the spectrum, approach or exceed the detector sensitivity threshold, indicating potential detectability of the corresponding zoom-whirl signatures.}\label{fig:GW-5}
\end{figure}

Figure~\ref{fig:GW-2} displays the Fourier amplitudes $|\tilde{\mathit{h}}_{+,\times}(f)|$ associated with the periodic EMRI waveforms. The spectra exhibit a discrete comb-like structure, with power concentrated in a sequence of narrow harmonic peaks generated by the underlying zoom-whirl motion. This pattern reflects the periodic character of the orbit and encodes the corresponding orbital frequencies. The distribution of spectral power depends on the orbit family as well as on the quintessence parameter $c_q$. As $c_q$ varies, the relative height of the harmonic peaks and the high-frequency tail of the spectrum are modified, indicating that the quintessence background changes the strong-field orbital dynamics and, hence, the spectral content of the emitted radiation.
For visual clarity, the numerically computed characteristic strain $\mathit{h}_{c}(f)$ defined in Eq.~\eqref{Eq:GW-9} is smoothed using a running average over 30 frequency bins. Figure~\ref{fig:GW-5} compares the resulting strain spectra for representative periodic orbits with the LISA sensitivity curve. The signal power is concentrated in the millihertz band, and the high-frequency part of the spectrum approaches or partially exceeds the LISA sensitivity threshold over a limited frequency range. This indicates that the strongest zoom-whirl harmonics generated by periodic EMRI motion in the present BH background may fall within the observational window of future space-based detectors~\cite{Ahmed:2025azu,Gair:2017ynp,Babak:2017tow}.

From a physical perspective, Fig.~\ref{fig:GW-5} shows how the strong-field geodesic structure of the orbit is transferred into an observable frequency-domain signature. The characteristic strain inherits the harmonic content of the underlying periodic motion, while the location and amplitude of the spectral features are controlled by the BH geometry through its effect on the orbital dynamics. Consequently, variations in the background parameters modify not only the time-domain waveform and Fourier amplitudes but also the portion of the signal that enters the detector sensitivity band. This makes the characteristic strain a useful diagnostic for assessing the potential detectability of zoom-whirl signatures and for probing environmental deformations of the central BH spacetime.

In this sense, the characteristic strain provides the direct link between strong-field orbital dynamics and observational accessibility, translating the imprint of the background geometry on periodic geodesic motion into a measurable GW signature.

Before closing, it is important to emphasize that the waveform modifications induced by the quintessence strength parameter $c_q$ are not necessarily unique. Changes in the phase evolution, burst structure, and harmonic content of the signal can, in principle, be partially degenerate with variations in the spin of a Kerr BH or with other orbital parameters of the EMRI system. Accordingly, the present results should not be interpreted as establishing $c_q$ as an unambiguous standalone probe of strong-field quintessence effects. Rather, they show that the quintessence background can generate systematic deformations in both the time and frequency-domain GW observables. Determining whether such imprints can be distinguished from Kerr spin and other astrophysical degeneracies requires a dedicated Bayesian parameter-estimation analysis with more complete waveform models, which lies beyond the scope of the present work.
\section{Conclusion and Discussion} \label{sec:con}

In this work, we investigated the timelike geodesic structure and GW signatures of periodic motion around a static magnetically charged BH immersed in a quintessence background. Our aim was to understand how the background parameters of the spacetime influence bound motion, the taxonomy of periodic timelike orbits, and the gravitational radiation emitted by an EMRI source moving in such a geometry. Since EMRIs encode the strong-field properties of the central compact object through the orbital dynamics of the smaller body, they provide a natural framework for probing nonvacuum BH spacetimes with future space-based detectors.

We first analyzed the effective potential governing timelike motion and determined the allowed region of bound orbits in terms of the conserved energy and angular momentum. In particular, we studied the MBO and the ISCO, which delimit the parameter range relevant for periodic timelike motion. Our results show that the quintessence background modifies the structure of the effective potential and shifts the corresponding $\mathit{L}$-$\mathit{E}$ domain of bound motion, thereby changing the location of turning points and the orbital conditions under which periodic trajectories can exist.

We then classified the resulting bound periodic orbits using the zoom-whirl taxonomy labeled by the triplet $(\mathit{z},\mathit{w},\mathit{v})$. The analysis shows that the quintessence parameter $c_q$ deforms the geometry of the trajectories while preserving their topological class, and systematically alters the associated orbital scales and turning-point structure. In this sense, the background quintessence field does not merely perturb individual geodesics, but reshapes the organization of the periodic orbit sector itself.
Tables~\ref{tab:ISCO_IBCO_lambda}, \ref{tab:energy_levels_cq}, and \ref{tab:angular_levels_cq} provide the corresponding quantitative support, showing that the characteristic radii, angular-momentum thresholds, and the conserved quantities associated with representative periodic families all vary systematically with $c_q$. Hence, the quintessence background affects not only the geometry of the bound trajectories but also the energetic conditions required for periodic zoom-whirl motion.

To explore the observational consequences of these effects, we modeled the gravitational radiation emitted by periodic EMRI orbits within the numerical kludge approximation. The resulting waveforms exhibit the characteristic burst-like structure of zoom-whirl motion: low-amplitude segments are associated with the zoom phase, whereas the whirl motion near periapsis generates short intervals of enhanced amplitude and frequency. We found that varying $c_q$ produces visible changes in the phase evolution, burst timing, and spectral content of the signal, while the qualitative waveform morphology remains tied to the underlying periodic orbit family. The corresponding Fourier spectra display a discrete harmonic structure, reflecting the periodic nature of the geodesic motion and its imprint on the emitted radiation.

We also examined the characteristic strain of the resulting signals and compared it with the LISA sensitivity curve. For the representative EMRI configurations considered here, the strongest harmonic components lie in the millihertz band and portions of the signal approach, or partially enter, the sensitivity window of future space-based detectors. This suggests that periodic zoom-whirl signatures generated in magnetically charged BH spacetimes with quintessence may provide potentially observable strong-field features in EMRI gravitational-wave signals.

At the same time, it is important to emphasize that the waveform deformations induced by the quintessence parameter $c_q$ are not necessarily unique. Similar changes in the phase evolution, burst structure, and harmonic content may in principle be partially degenerate with variations in Kerr spin or with other orbital parameters of the inspiralling system. The present analysis therefore should not be interpreted as establishing $c_q$ as an unambiguous standalone probe of strong-field quintessence effects. Rather, it shows that a quintessence background can leave systematic imprints on periodic orbit dynamics and on the associated time and frequency-domain GW observables. A full assessment of parameter degeneracies and detectability will require more complete waveform modeling and dedicated Bayesian parameter-estimation analyses, which we leave for future work.

\section*{Acknowledgments}
The author is sincerely grateful to Prof.~Xiao-Mei Kuang and Dr.~Yong-Zhuang Li for many valuable discussions, insightful suggestions, continuous guidance, and constructive comments that significantly improved this study.

\bibliographystyle{utphys}
\bibliography{ref}
\end{document}